\documentclass[12pt]{article}
\pdfoutput=1

\usepackage{jcappub}

\usepackage{amsthm,graphicx,multirow}
\usepackage{epsfig}
\usepackage{latexsym, amssymb} 
\usepackage{amsmath}
\usepackage{epstopdf}
\usepackage{hyperref}

\newcommand{\dd}{\mathrm{d}}

\usepackage[normalem]{ulem}

\begin{document}
\title{Non-linear damping of superimposed primordial oscillations on the matter power spectrum in galaxy surveys}

\author[1,2,3,4]{Mario Ballardini,}
\author[5,6,7,10]{Riccardo Murgia,}
\author[1,8,3]{Marco Baldi,}
\author[2,3]{Fabio Finelli,}
\author[5,6,7,9]{Matteo Viel}
\date{\today}

\affiliation[1]{Dipartimento di Fisica e Astronomia, Alma Mater Studiorum Universit\`a di Bologna, via Gobetti93/2, I-40129 Bologna, Italy}
\affiliation[2]{INAF/OAS Bologna, via Piero Gobetti 101, I-40129 Bologna, Italy}
\affiliation[3]{INFN, Sezione di Bologna, via Irnerio 46, I-40126 Bologna, Italy}
\affiliation[4]{Department of Physics \& Astronomy, University of the Western Cape, Modderdam Road P/Bag X17, Bellville 7530, South Africa}
\affiliation[5]{SISSA, via Bonomea 265, 34136 Trieste, Italy}
\affiliation[6]{INFN, Sezione di Trieste, via Valerio 2, 34127 Trieste, Italy}
\affiliation[7]{IFPU, Institute for Fundamental Physics of the Universe, via Beirut 2, 34151 Trieste, Italy}
\affiliation[8]{INAF/OAS Bologna, via Piero Gobetti 93/3, I-40129 Bologna, Italy}
\affiliation[9]{INAF, Osservatorio Astronomico di Trieste, via Tiepolo 11, I-34131 Trieste, Italy}
\affiliation[10]{Laboratoire Univers $\&$ Particules Montpellier (LUPM), CNRS $\&$ Universit\`{e} de Montpellier (UMR-5299), Place Eug\`{e}ne Bataillon, F-34095 Montpellier Cedex 05, France}

\emailAdd{mario.ballardini@inaf.it}
\emailAdd{riccardo.murgia@umontpellier.fr}
\emailAdd{marco.baldi5@unibo.it}
\emailAdd{fabio.finelli@inaf.it}
\emailAdd{viel@sissa.it}

\abstract{Galaxy surveys are an important probe for superimposed oscillations on the 
primordial power spectrum of curvature perturbations, which are predicted in several 
theoretical models of inflation and its alternatives. In order to exploit the full 
cosmological information in galaxy surveys it is necessary to study the matter power 
spectrum to fully non-linear scales.
We therefore study the non-linear clustering in models with superimposed linear and 
logarithmic oscillations to the primordial power spectrum by running high-resolution 
dark-matter-only N-body simulations.
We fit a Gaussian envelope for the non-linear damping of 
superimposed oscillations in the matter power spectrum to the results of the N-body simulations 
for $k \lesssim 0.6\ h/$Mpc at $0 \leq z \leq 5$ with an accuracy below the percent.
We finally use this fitting formula to forecast the capabilities of future galaxy surveys, 
such as Euclid and Subaru, to probe primordial oscillation down to non-linear scales 
alone and in combination with the information contained in CMB anisotropies.}

\maketitle

\section{Introduction}
The study of departures from a simple power-law in the primordial power spectrum (PPS) 
of curvature perturbations has been fueled by theoretical advances and observational 
progress over many years (see Ref.~\cite{Chluba:2015bqa} for a review).

From a theoretical perspective, departures from a simple power-law can be a signature of 
the breakdown of any of the assumptions behind standard single field slow-roll inflation 
with Bunch-Davies initial conditions for quantum fluctuations 
\cite{Starobinsky:1980te,Guth:1980zm,Linde:1981mu,Linde:1983gd,Albrecht:1982wi}.
In the analogy in which primordial fluctuations can be seen as a cosmological collider for 
the physics of the early Universe \cite{Chen:2009zp,Arkani-Hamed:2015bza}, these features 
could help in discriminating  between inflation and alternative scenarios, or could provide 
hints for inflaton dynamics beyond slow-roll, and new heavy particles.

From the observational side, departures from a simple power-law in the PPS are of extreme 
interest, despite the tighter and tighter constraints on their size due to the increasing 
precision of cosmological observations. Well motivated theoretical models with features in 
the PPS have led to an improvement in the fit to cosmic microwave background (CMB) anisotropies 
data with respect to the simplest power-law spectrum since the WMAP first year data 
\cite{Peiris:2003ff} to the final {\em Planck} legacy data release \cite{Akrami:2018odb}. 
However, these improvements in the fit come at the expense of having extra parameters and 
these models have not been preferred over the simplest power-law spectrum at a statistically 
significant level so far, see e.g. \cite{Akrami:2018odb}.

Next generation of cosmological observations will help in explaining whether the hints for 
departures from a power-law spectrum for primordial fluctuations have a physical origin or 
are a mere statistical fluctuation.
In particular, large-scale structure (LSS) surveys are very promising 
\cite{Zhan:2005rz,Huang:2012mr,Hu:2014hra,Chen:2016zuu,Chen:2016vvw,Ballardini:2016hpi,Xu:2016kwz,Fard:2017oex,Palma:2017wxu,LHuillier:2017lgm,Ballardini:2017qwq,Ballardini:2018noo,Zeng:2018ufm} since they can probe the PPS to smaller scales and can increase the range of scales which are independently scanned 
by CMB anisotropy measurements.
It has been already quantified how future LSS surveys could significantly improve current 
constraints on theoretical models with features in the PPS by just using linear scales, 
i.e. $k \lesssim 0.1\ h/$Mpc \cite{Huang:2012mr,Chen:2016vvw,Ballardini:2016hpi,Ballardini:2017qwq}.
On the other hand, non-linear effects are important on most of the scales which are probed 
by LSS surveys and for these models are not accurately described by {\texttt {Halofit}}. 
Therefore, the study of non-linear dynamic is required to have full access to the 
information for primordial features contained in the dark matter (DM) power spectrum 
measurements as already studied in Refs.~\cite{Vlah:2015zda,Beutler:2019ojk,Vasudevan:2019ewf}. 

In this paper, we study two templates for the PPS which include undamped oscillations at small 
scales and therefore need the understanding of the non-linear gravitational instability at 
the scales of interest for galaxy surveys. Among the several models of primordial features, 
we choose the case of undamped linear or logarithmic oscillations superimposed 
on all the scales of interest to a PPS described by a power-law 
\cite{Chen:2008wn,Wang:2002hf,Bean:2008na,Flauger:2009ab,Chen:2011zf,Chen:2011tu,Danielsson:2002kx,Easther:2002xe,Martin:2003kp}: these are the theoretical models which lead to the largest improvement in 
the fit of CMB anisotropies and which need the understanding of non-linear clustering on scales 
$k \gtrsim 0.1\ h/$Mpc, given the presence of oscillations on all the scales.
With these linear and logarithmic oscillations superimposed to the PPS we run a set of 
high-resolution DM-only cosmological simulations with 1,024$^3$ DM particles in a comoving box 
with side length of 1,024 Mpc$/h$ (see \cite{LHuillier:2017lgm} for N-body simulations with different 
type of primordial features).
We then develop a fitting function calibrated against a set of N-body simulations with features 
in the PPS, following the approach previously used in {\texttt {Halofit}} 
\cite{Smith:2002dz,Takahashi:2012em} or {\texttt {HM-Code}} \cite{Mead:2015yca,Mead:2016zqy} 
for $\Lambda$CDM and some of its extensions.

The paper is organized as follows. We begin in Sec.~\ref{sec:models} introducing the two 
templates for oscillatory features that we study. In Sec.~\ref{sec:sims}, we describe the 
simulations. In Sec.~\ref{sec:fit}, we use a Gaussian envelope for the non-linear damping and 
we calibrate it against the N-body simulations; we also compare our findings with the 
leading-order theoretical predictions for the damping from Refs.~\cite{Beutler:2019ojk,Vasudevan:2019ewf}. 
We discuss in Sec.~\ref{sec:BAO} the damping of the baryon acoustic oscillations (BAO) 
features versus the damping of the primordial linear oscillations. We run a series of 
forecasts in Sec.~\ref{sec:analysis} with galaxy clustering up to $k_{\rm max} = 0.6\ h/$Mpc 
in combination with CMB for a Euclid-like experiment and Subaru Prime Focus Spectrograph (PFS), 
and we discuss the results in Sec.~\ref{sec:results}. Sec.~\ref{sec:conclusion} contains 
our conclusions.

\section{Superimposed oscillations on the primordial power spectrum} \label{sec:models}

The type of superimposed oscillatory features on the PPS which we study in this paper are 
predicted in several well motivated theoretical models. These features can be generated by 
an oscillatory signal in time in the inflationary field potential or in the internal field 
space with a frequency larger than the Hubble parameter $H$ able to resonate with the curvature 
modes inside the horizon \cite{Chen:2008wn}.
They can be realized in many contexts as in axion inflation \cite{Wang:2002hf}, 
 small-field models such as brane inflation \cite{Bean:2008na}, 
 large-field models in string theory such as axion monodromy \cite{Flauger:2009ab},
or as oscillations of massive fields \cite{Chen:2011zf,Chen:2011tu}.
Superimposed oscillations on the PPS are also generated when inflaton temporarily deviates from 
the attractor solution at some point during its evolution \cite{Starobinsky:1992ts,Adams:2001vc}
or for non Bunch-Davies initial conditions \cite{Danielsson:2002kx,Easther:2002xe,Martin:2003kp}.

We study two templates with superimposed oscillations on the PPS \cite{Chen:2008wn}, the first 
with linear oscillations
\begin{equation} \label{eqn:pk_lin}
    P_\zeta (k) = P_{\zeta,0} (k) \left[ 1 + {\cal A}_{\rm lin} \cos\left(\omega_{\rm lin}\frac{k}{k_*}+\phi_{\rm lin}\right) \right] \,,
\end{equation}
and the second with logarithmic oscillations
\begin{equation} \label{eqn:pk_log}
    P_\zeta (k) = P_{\zeta,0} (k) \left[ 1 + {\cal A}_{\rm log} \cos\left(\omega_{\rm log}\log \frac{k}{k_*}+\phi_{\rm log}\right) \right] \,,
\end{equation}
where $P_{\zeta,0} (k) = A_s (k/k_*)^{n_s - 1}$ is the standard power-law PPS with pivot scale 
$k_* = 0.05$ Mpc$^{-1}$.

\section{Accurate fitting formula for the non-linear matter power spectrum with superimposed primordial oscillations}

Galaxies trace the invisible cold dark matter (CDM) distribution and we can estimate 
their power spectrum to extract information on the underlying power spectrum of primordial fluctuations.
While on linear scales the matter power spectrum can be computed for any given initial conditions 
and cosmological model with dedicated Einstein-Boltzmann solvers like 
{\tt CAMB}\footnote{\href{https://github.com/cmbant/CAMB}{https://github.com/cmbant/CAMB}}~\cite{Lewis:1999bs,Howlett:2012mh} 
or {\tt CLASS}\footnote{\href{https://github.com/lesgourg/class_public}{https://github.com/lesgourg/class\_public}}~\cite{Lesgourgues:2011re,Blas:2011rf}, 
in the non-linear regime, one has to rely on cosmological N-body simulations to study the non-linear 
gravitational evolution for every extension of the $\Lambda$CDM cosmological model.

The halofit model has been successfully used to predict the small-scale non-linearities for the 
$\Lambda$CDM cosmology and some of its simplest extensions such as models including massive neutrinos 
\cite{Bird:2011rb} or non-standard dark energy equations of state (wCDM) \cite{Takahashi:2012em}. 
So far, this programme has not yet been pursued for models with primordial superimposed oscillations, 
and we aim to start the process with the present analysis. In particular, we wonder how superimposed 
oscillations will be damped on non-linear scales and if there will be any additional effect like a 
running of the frequency or a de-phasing of the oscillations due to the non-linear evolution of the 
perturbations.

\subsection{Cosmological simulations} \label{sec:sims}

\begin{table}
\setlength{\tabcolsep}{5pt}\renewcommand{\arraystretch}{1.2}
\centering
\begin{tabular}{|c|c|c|c|}
\hline
Model & ${\cal A}$ & $\log_{10}\omega$ & $\phi/(2\pi)$ \\
\hline  
Lin. Osc. & 0.03 &   0.8  &  0.0  \\
Lin. Osc. & 0.03 &   0.8  &  0.6  \\
Lin. Osc. & 0.03 &   0.87 &  0.0  \\
Lin. Osc. & 0.03 &   0.87 &  0.2  \\
Lin. Osc. & 0.03 &   1.0  &  0.4  \\
Lin. Osc. & 0.03 &   1.0  &  0.6  \\
Log. Osc. & 0.03 &   0.8  &  0.2  \\
Log. Osc. & 0.03 &   0.87 &  0.4  \\
Log. Osc. & 0.03 &   1.26 &  0.8  \\
Log. Osc. & 0.03 &   1.5  &  0.6  \\
\hline
\end{tabular}
\caption{Here we report the 10 cosmological models that we have considered for our analyses, each of them identified by an amplitude ${\cal A}$, a frequency $\omega$, and a phase $\phi/(2\pi)$ (see Eqs.~\eqref{eqn:pk_lin}~and~\eqref{eqn:pk_log}). The first six (last four) models correspond to a superimposed linear (logarithmic) oscillation pattern.\label{tab:osc_models}}
\end{table}

In order to perform our analysis, we have run a set of 10+1 high-resolution DM-only cosmological simulations corresponding to 6 (4) models with superimposed linear (logarithmic) oscillations, all of them listed in Tab.~\ref{tab:osc_models}, plus the standard $\Lambda$CDM case. Each of the simulations follows the non-linear evolution of 1,024$^3$ DM particles in a comoving box with side length of 1,024 Mpc$/h$, using a gravitational softening length of 25 kpc$/h$, down to redshift $z=0$. The cosmological parameters have been fixed to the following values: $\Omega_m = 0.321$, $\Omega_\Lambda = 0.679$, $n_s = 0.963$, $H_0 = 66.9$~km~s$^{-1}$~Mpc$^{-1}$, and $\sigma_8 = 0.8$. To minimise the noise induced by cosmic variance, we also performed 3 more simulations with larger boxes, i.e. 1,024$^3$ DM particles in a 2,048 Mpc$/h$ size length box,  only for the highest-frequency logarithmic models, which are the most 
sensitive to the box size, and the $\Lambda$CDM case.

\begin{figure}
\centering
\includegraphics[width=1.1\textwidth]{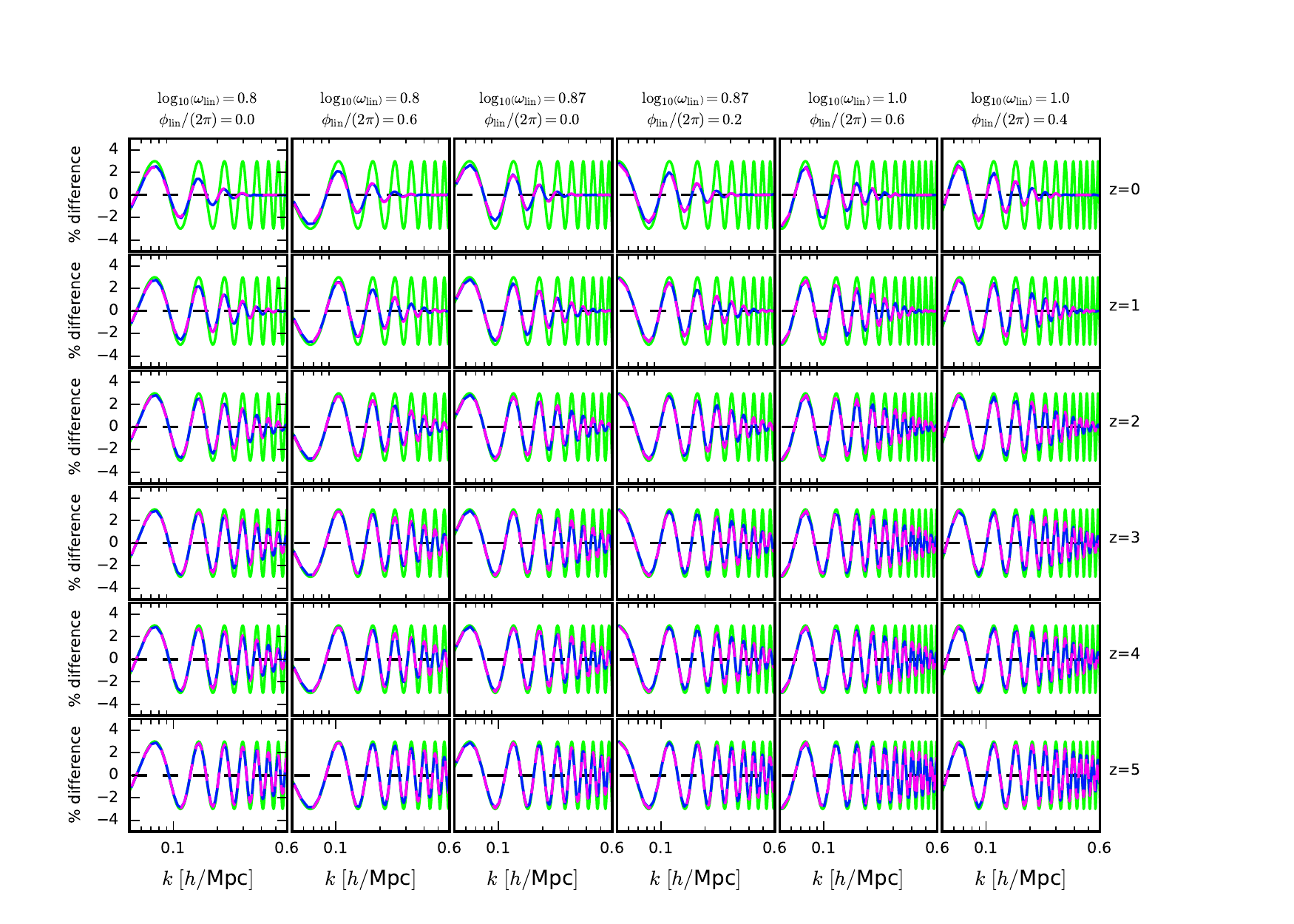}
\caption{Relative differences with respect to the $\Lambda$CDM matter power spectrum between 
non-linear matter power spectrum with undamped superimposed oscillation (green), non-linear 
matter power spectrum obtained from the simulations (blue), and non-linear matter power spectrum 
reconstructed with our semi-analytical damping \eqref{eqn:pk_fit} (dashed magenta) for the 
template with linear oscillation \eqref{eqn:pk_lin} at five different redshift $z=0, 1, 2, 3, 4, 5$ 
from the top to the bottom respectively.}
\label{fig:LINfit}
\end{figure}

\begin{figure}
\centering
\includegraphics[width=1.1\textwidth]{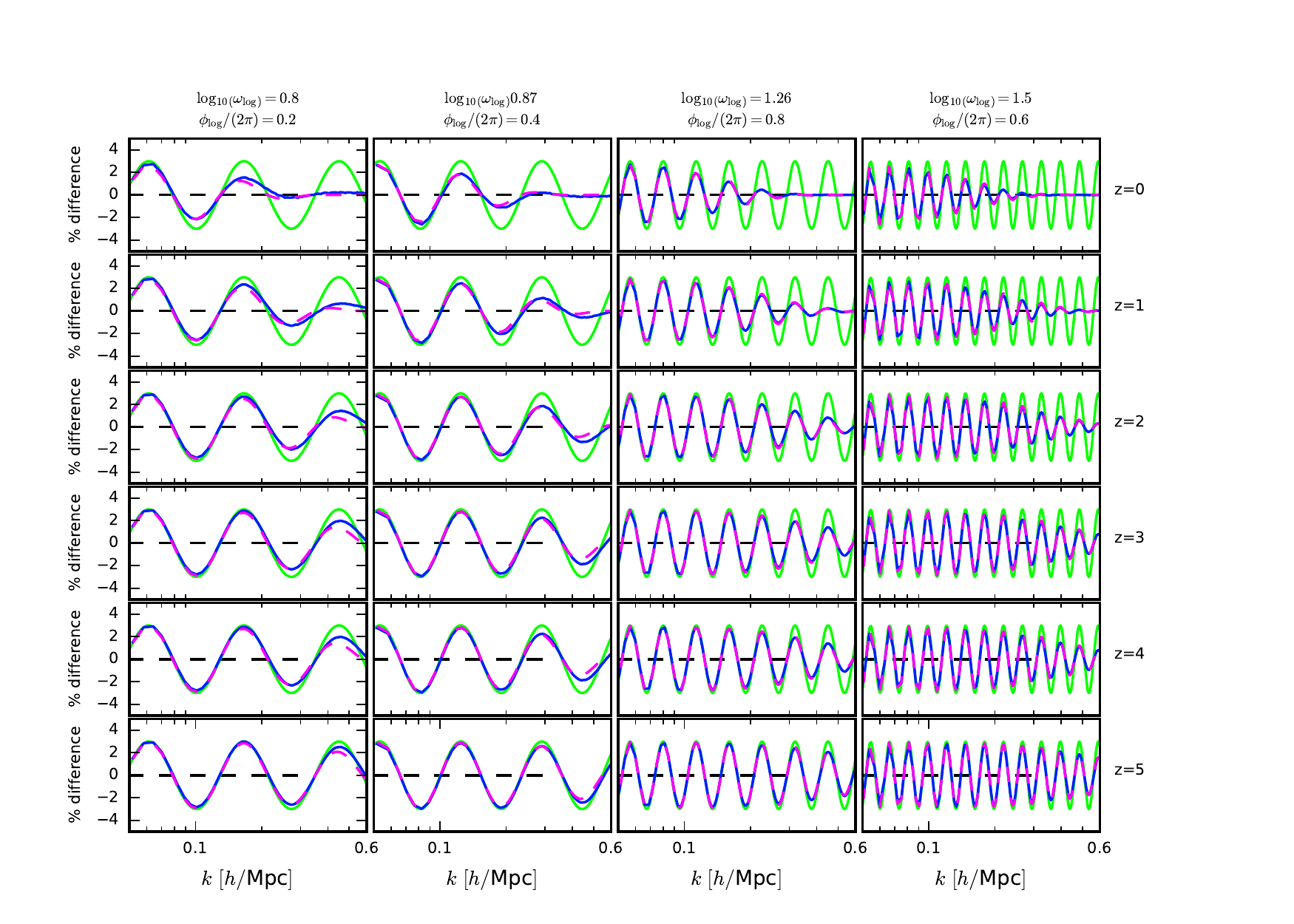}
\caption{Relative differences with respect to the $\Lambda$CDM matter power spectrum between 
non-linear matter power spectrum with undamped superimposed oscillation (green), non-linear 
matter power spectrum obtained from the simulations (blue), and non-linear matter power spectrum 
reconstructed with our semi-analytical damping \eqref{eqn:pk_fit} (dashed magenta) for the 
template with logarithmic oscillation \eqref{eqn:pk_log} at five different redshift $z=0,1,2,3,4,5$ 
from the top to the bottom respectively.}
\label{fig:LOGfit}
\end{figure}

All simulations have been run with the N-body code {\texttt{GADGET-3}}, a modified version of the publicly available numerical code {\texttt{GADGET-2}}~\cite{Springel:2000qu,Springel:2005mi}. The initial conditions have been produced by displacing the DM particles from a cubic Cartesian grid according to second-order Lagrangian Perturbation Theory, with the {\texttt{2LPTic}} code~\cite{Crocce:2006ve}, at redshift $z=99$. The corresponding input linear matter power spectra were computed with a modified version of the publicly available code {\texttt{CAMB}}, with the superimposed oscillations given by Eq.~\eqref{eqn:pk_lin} for the linear cases, and by Eq.~\eqref{eqn:pk_log} for logarithmic cases. The values assigned to the amplitude ${\cal A}$, the frequency $\omega$, and the normalized phase $\phi/(2\pi)$, associated with each of the models are reported in Tab.~\ref{tab:osc_models}.
In generating the initial conditions we turned off the Rayleigh sampling as done in Ref.~\cite{Viel:2010bn}, in order to fix the mode amplitude to the expected value of the linear power spectrum. We explicitly check that this aspect does not bias any of our results that are always cast in terms of ratios between the 
case including primordial oscillations and the corresponding baseline power-law case.
For a more comprehensive analysis of Rayleigh sampling and paired fixed field simulation we refer to Ref.~\cite{Villaescusa-Navarro:2018bpd}.
On top of these simulations, we have used a Friends-of-Friends (FoF) algorithm~\cite{Davis:1985rj} with the standard linking length $b=0.2$, in order to identify particle groups and to extract the statistics of the associated DM halos. 

\begin{figure}
\centering
\includegraphics[width=.48\textwidth]{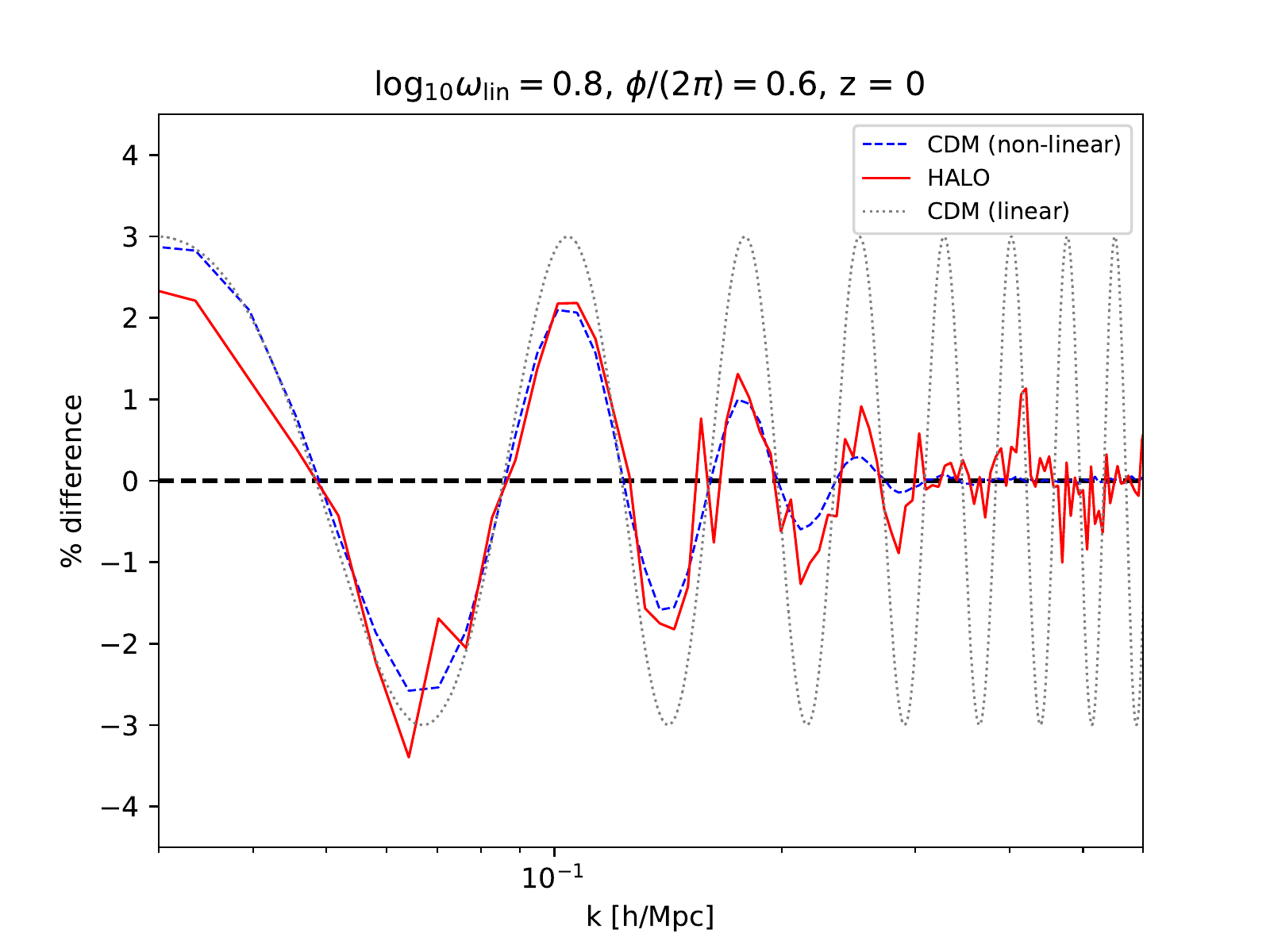}
\includegraphics[width=.48\textwidth]{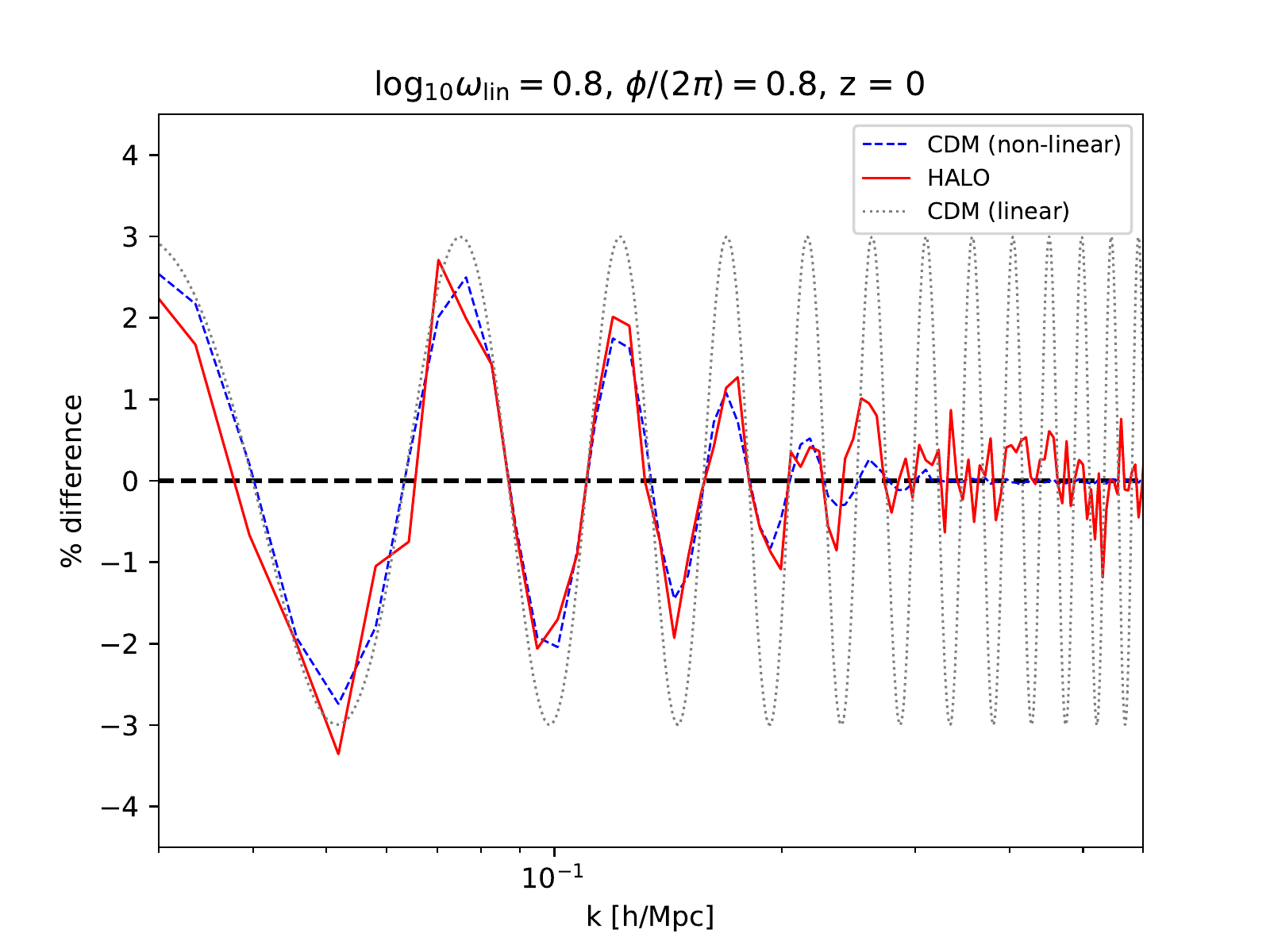}
\includegraphics[width=.48\textwidth]{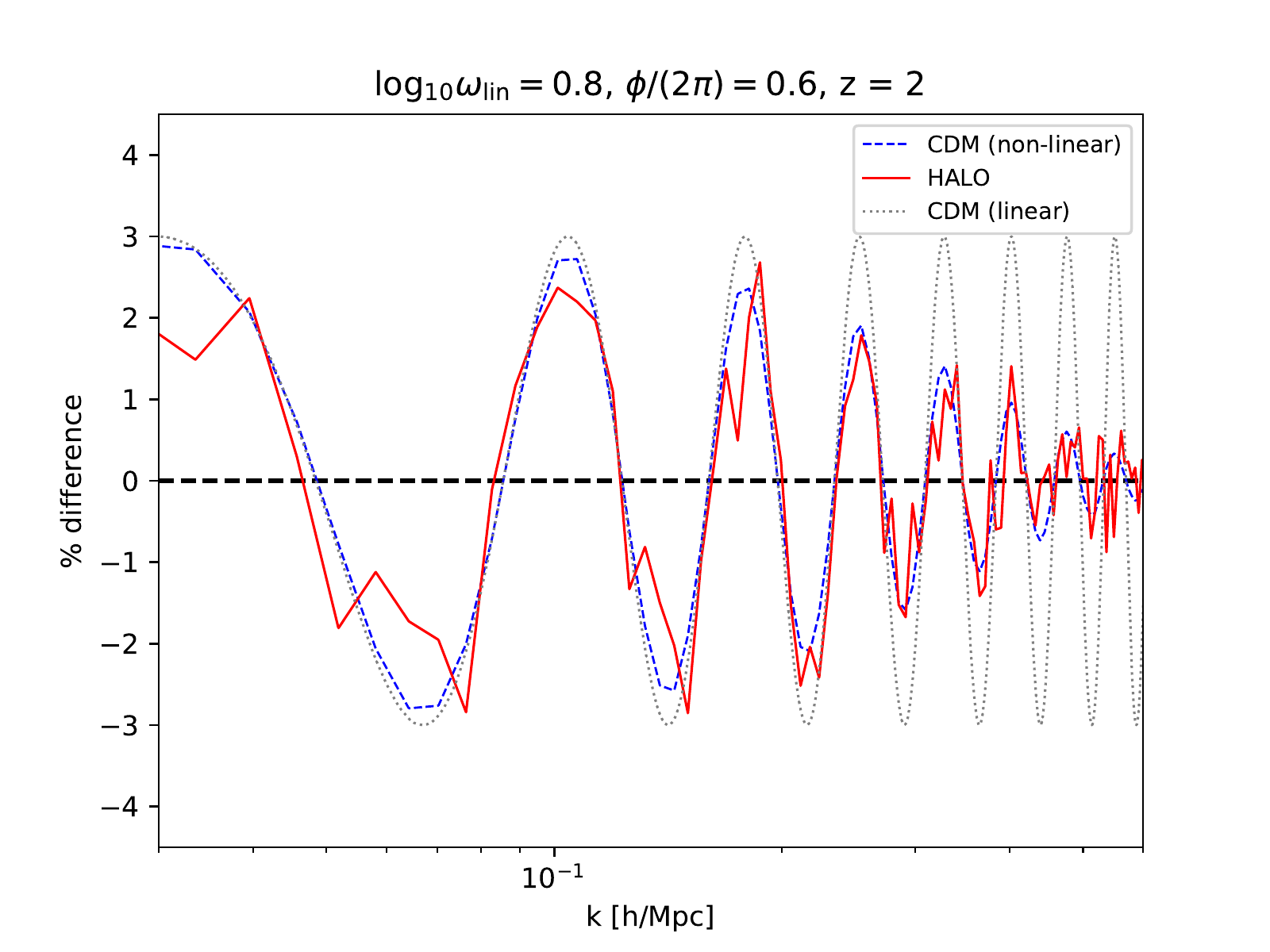}
\includegraphics[width=.48\textwidth]{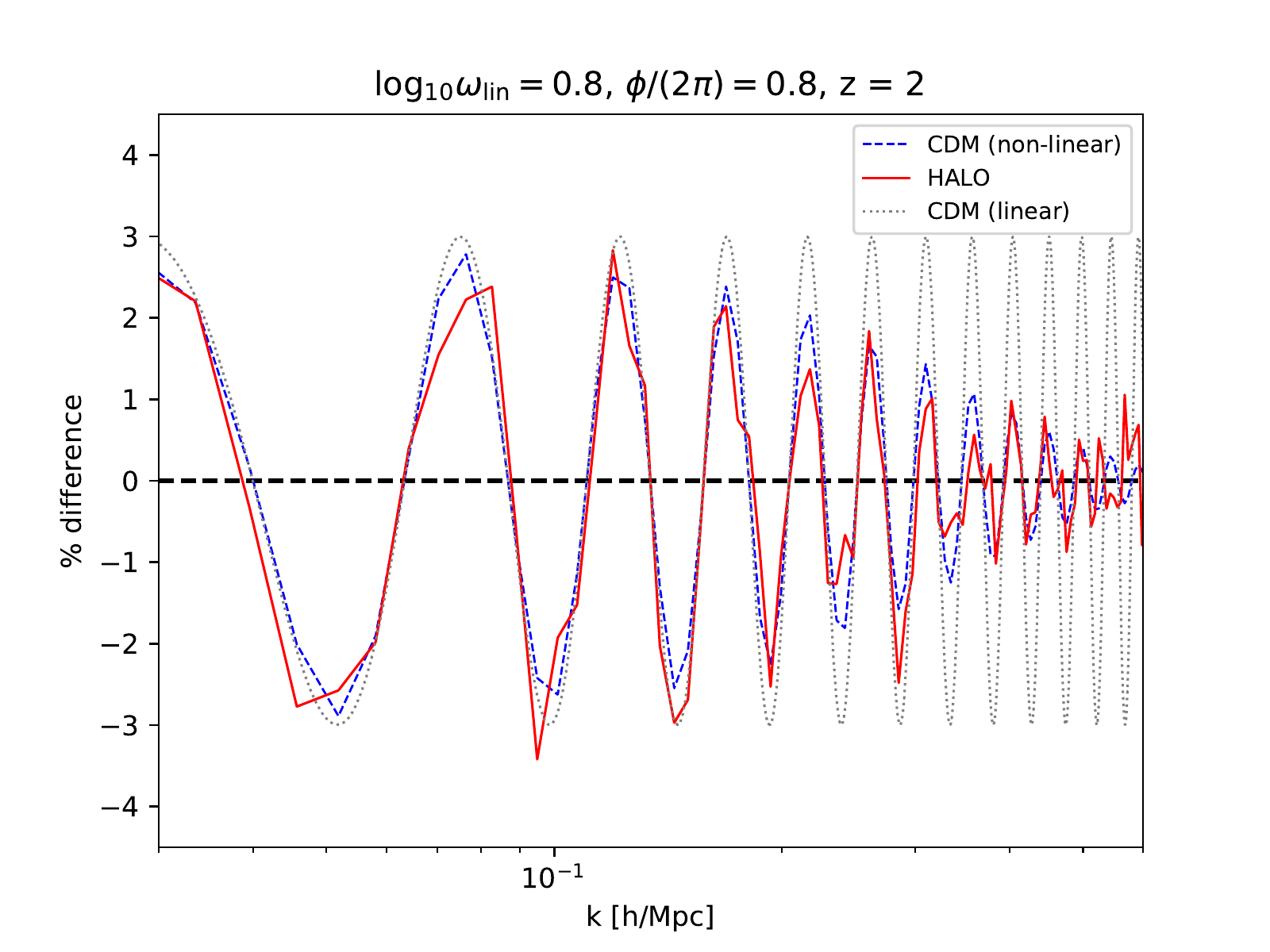}
\caption{Relative differences with respect to the $\Lambda$CDM case, for the DM halo power 
spectra computed from N-body simulations (red solid lines), for the template with linear 
oscillation \eqref{eqn:pk_lin} at two different redshift $z=0,2$, from the top to the bottom 
respectively. We show on the left the results for ${\cal A}_{\rm lin}=0.03$, 
$\log_{10} \omega_{\rm lin} = 0.8$, $\phi/(2\pi)=0.6$ and on the right for ${\cal A}_{\rm lin}=0.03$, 
$\log_{10} \omega_{\rm lin} = 1.0$, $\phi/(2\pi)=0.4$. As a reference, we also plot the 
corresponding linear (dotted gray lines) and non-linear (blue dashed lines) matter power spectra.}
\label{fig:LIN_pk_h}
\end{figure}

\begin{figure}
\centering
\includegraphics[width=.48\textwidth]{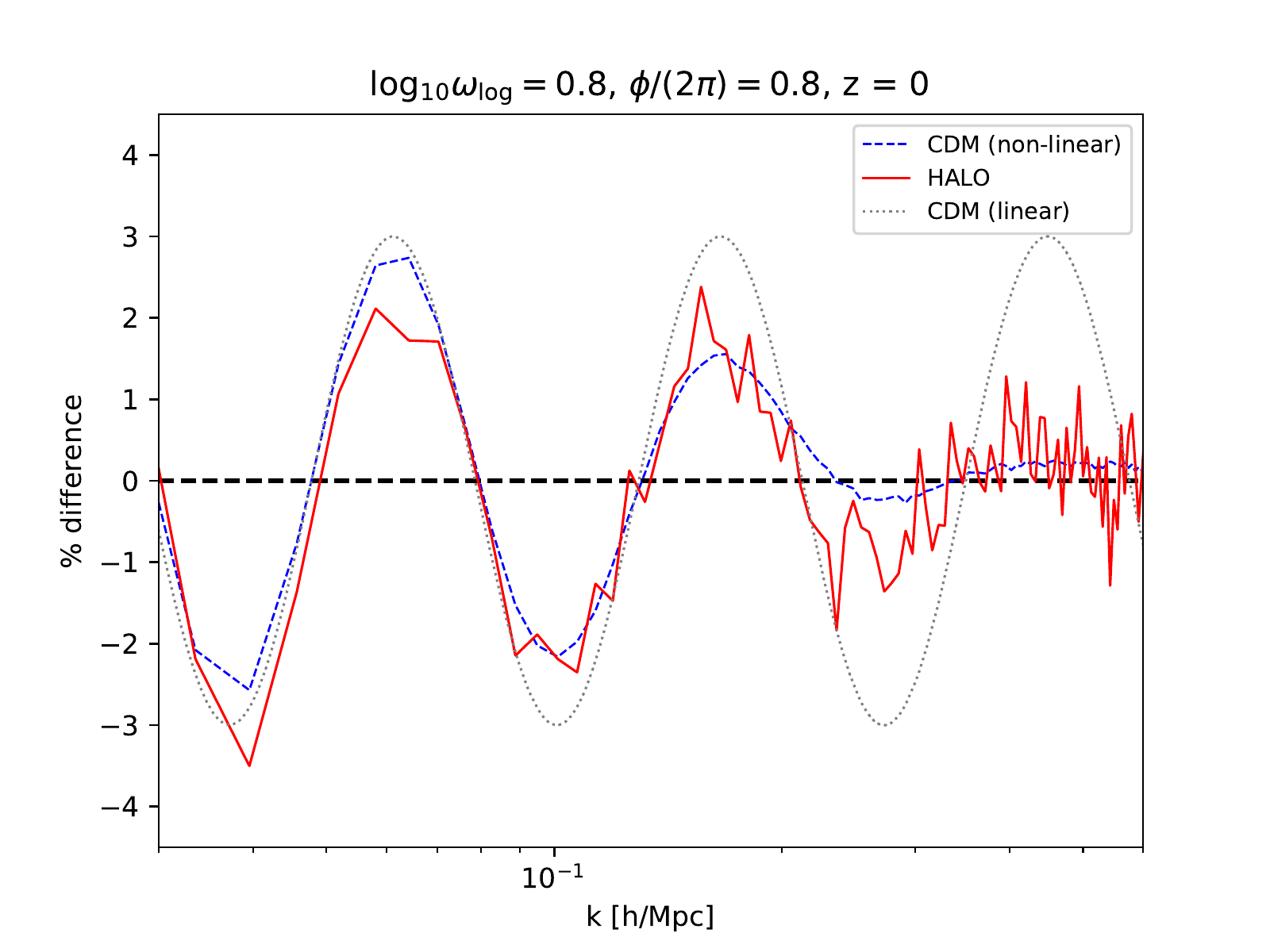}
\includegraphics[width=.48\textwidth]{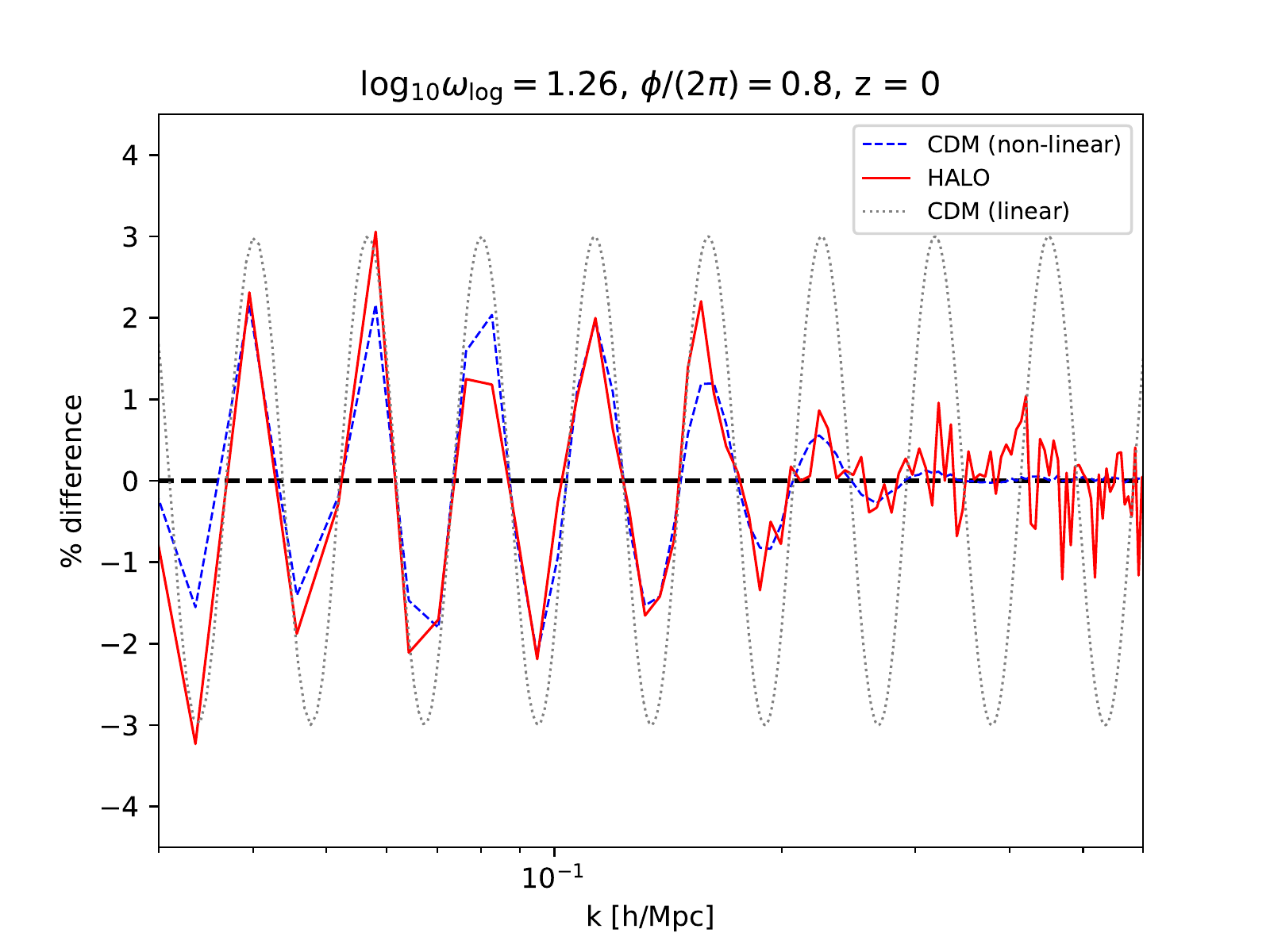}
\includegraphics[width=.48\textwidth]{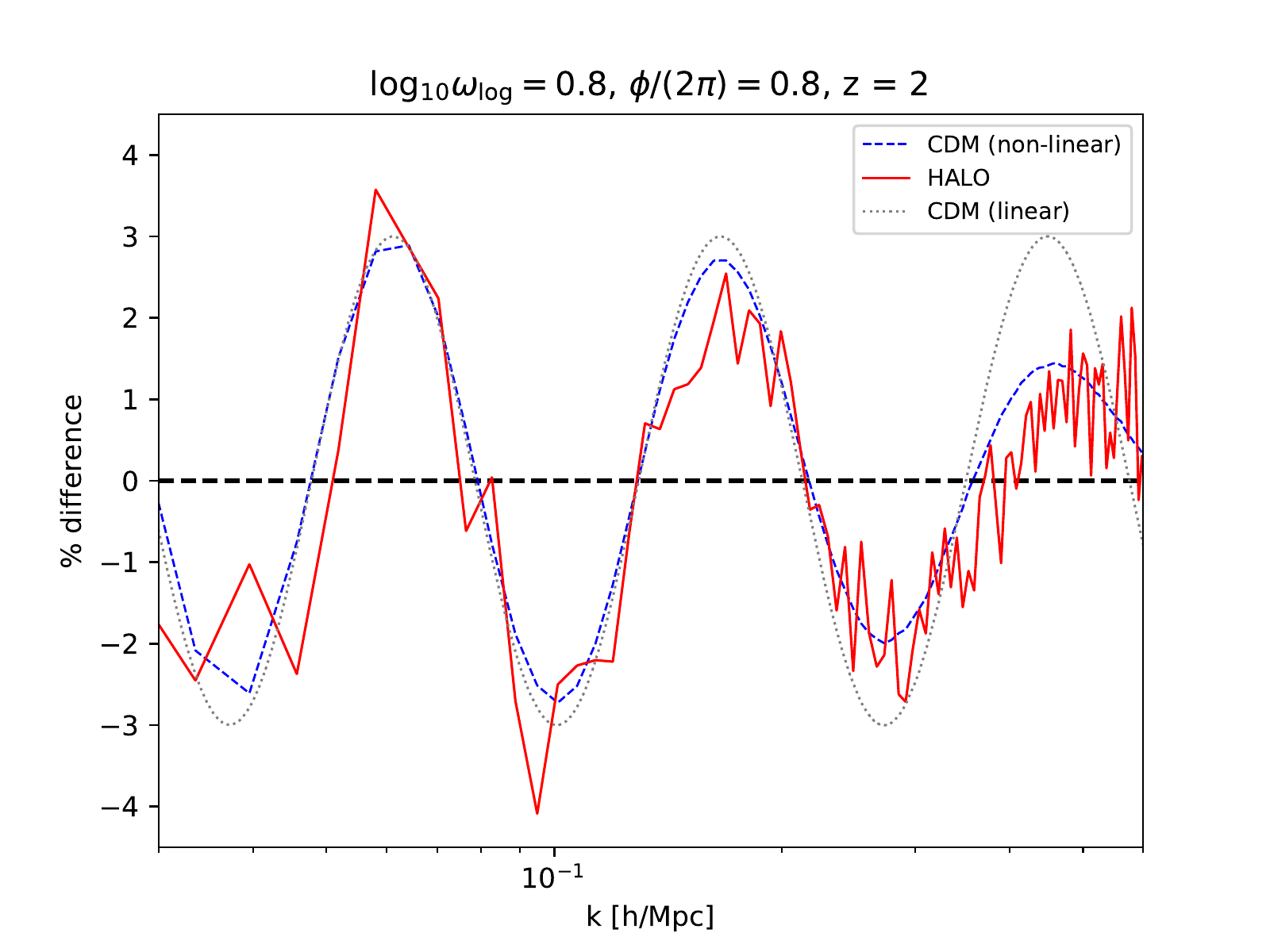}
\includegraphics[width=.48\textwidth]{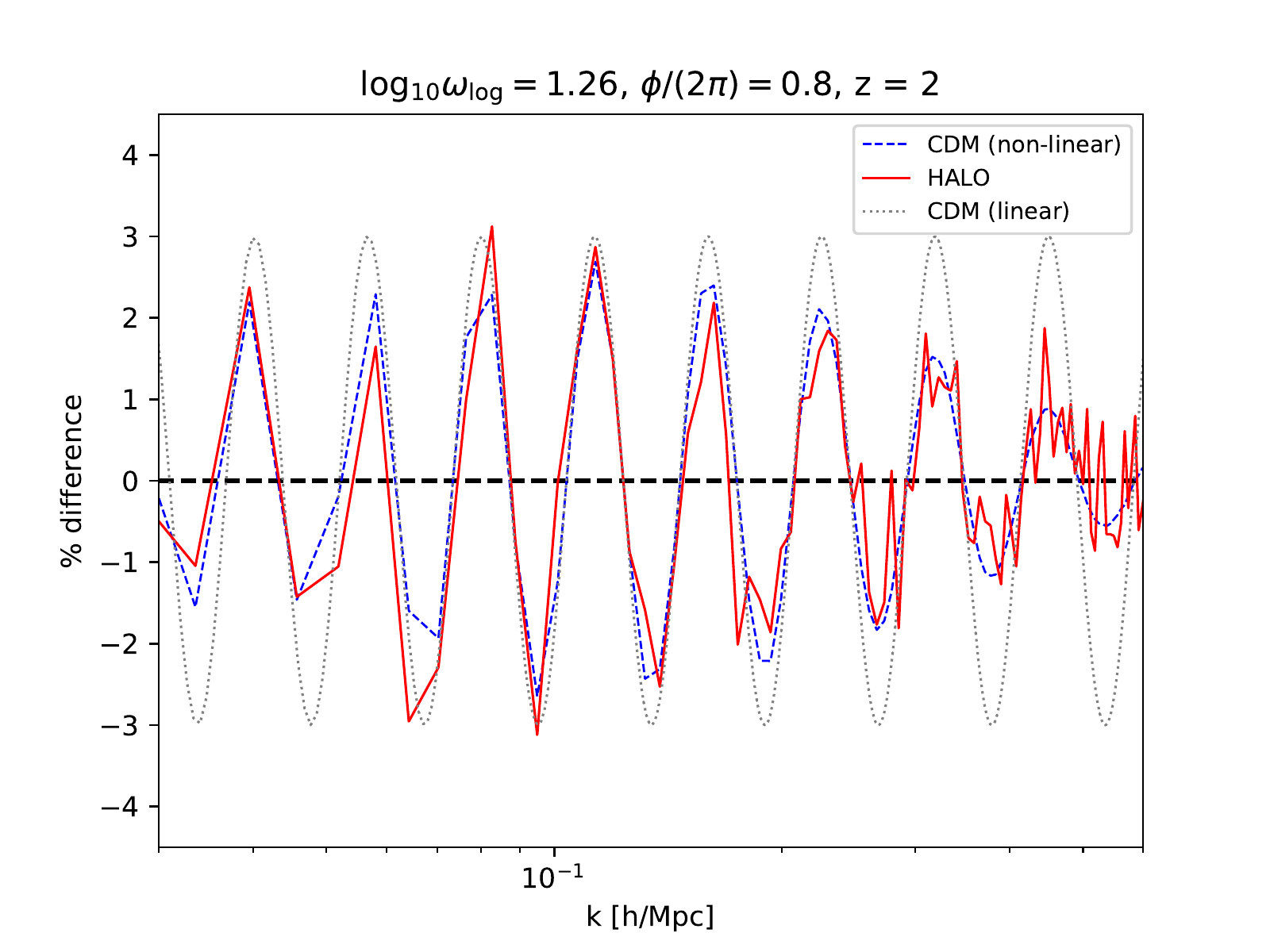}
\caption{\looseness=-1 Relative differences with respect to the $\Lambda$CDM case, for the 
DM halo power spectra computed from N-body simulations (red solid lines), for the template
with logarithmic oscillation \eqref{eqn:pk_log} at two different redshift $z=0,2$, from the top 
to the bottom respectively. We show on the left the results for ${\cal A}_{\rm log}=0.03$, 
$\log_{10} \omega_{\rm log} = 0.8$, $\phi/(2\pi)=0.2$ and on the right for ${\cal A}_{\rm log}=0.03$, 
$\log_{10} \omega_{\rm log} = 1.26$, $\phi/(2\pi)=0.8$. As a reference, we also plot the corresponding 
linear (dotted gray lines) and non-linear (blue dashed lines) matter power spectra.}
\label{fig:LOG_pk_h}
\end{figure}

For all simulations we have extracted the matter and halo power spectra $P_{m}(k,z)$ and 
$P_{h}(k,z)$ as a function of the Fourier wavemode $k$ and of the redshift $z$ by assigning 
the mass of tracer DM particles and individual collapsed halos to a Cartesian grid with 
$1,024^3$ cells through a Cloud-In-Cell mass assignment scheme. 
The visual inspection of the ratio of each model's power spectrum to the reference $\Lambda$CDM 
scenario shows how the primordial pattern of oscillations can still be clearly observed at 
low redshifts, with a significant damping of the small-scale oscillations which we show between 
$z=5$ and $z=0$, for both the DM (see Figs.~\ref{fig:LINfit} and \ref{fig:LOGfit} below) and 
the halos (see Figs.~\ref{fig:LIN_pk_h} and \ref{fig:LOG_pk_h} below) distributions.

\begin{figure}
\centering
\includegraphics[width=.7\textwidth]{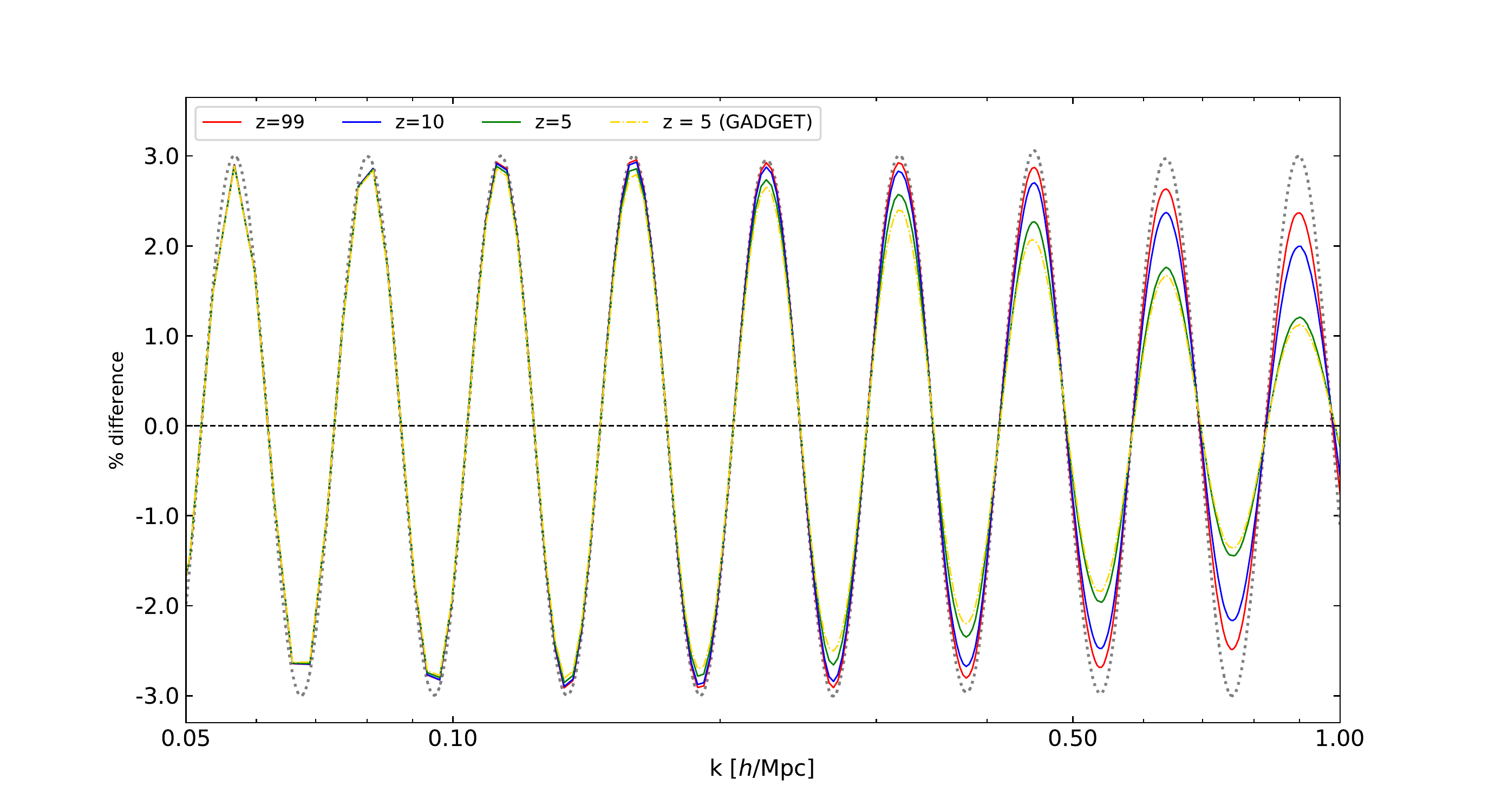} 
\caption{We show the relative differences between the non-linear matter power spectrum 
of one of the logarithmic models (${\cal A}_{\rm log}=0.03$, $\log_{10} \omega_{\rm log} = 0.8$, 
$\phi_{\rm log}/(2\pi)=0.2$) with respect to the $\Lambda$CDM case, as extracted from initial conditions 
produced through {\texttt{2LPTic}} (solid lines). We also show one power spectrum extracted from 
the corresponding N-body simulation, at $z=5$ (dot-dashed  line). As a reference, we also report the relative difference between the linear matter power spectra (gray dotted line).}
\label{fig:BAOcomparison1}
\end{figure}

In Fig.~\ref{fig:BAOcomparison1}, we show the relative differences 
between the non-linear matter power spectrum of one of the 
logarithmic models (${\cal A}_{\rm log}=0.03$, $\log_{10} \omega_{\rm log} = 0.8$, $\phi_{\rm log}/(2\pi)=0.2$) with 
respect to the $\Lambda$CDM case. The solid lines of Fig.~\ref{fig:BAOcomparison1} 
refer to spectra extracted from initial conditions produced through {\texttt{2LPTic}}, at different 
redshifts. For comparison, we also show one power spectrum extracted from the corresponding N-body 
simulation, at $z=5$, which is in very good agreement with the corresponding {\texttt{2LPTic}} output 
(dot-dashed line). The relative difference between the linear matter power spectra is plotted as a 
gray dotted line.

\subsection{Fit model} \label{sec:fit}

We use our simulations to calibrate the small-scale damping induced by non-linear 
dynamics on the oscillatory features superimposed in the matter power spectrum. To do so, 
we use a least chi-squared method to find the best-fit solution of
\begin{equation}
    \chi^2\left(\Sigma\right) = \sum_i \sum_{z=0}^{5} \sum_{k=k_{\rm min}}^{k_{\rm max}} 
    \left[\frac{P_{i{\rm ,fit}}(k,z,\Sigma)-P_{i{\rm ,sim}}(k,z)}{\sigma_i(k,z)}\right]^2 
\end{equation}
with linear loss function, where $i$ runs over the different best-fit for the features parameters 
listed in Tab.~\ref{tab:osc_models}, 
$P_{i{\rm ,fit}}(k,z)$ is our semi-analytic template to model non-linear effects for the 
superimposed oscillations, and $P_{i{\rm ,sim}}(k,z)$ is the non-linear matter 
spectrum from the simulations. 
We set the variance $\sigma_i(k,z)=P_{\rm sim}(k,z)$, where $P_{\rm sim}(k,z)$ is the non-linear 
matter power spectrum for a $\Lambda$CDM cosmology from the simulations.
We consider wavenumbers between $k_{\rm min}=0.05\ h/$Mpc and $k_{\rm max}=0.6\ h/$Mpc.

We write the semi-analytic template to model non-linear effects as:
\begin{equation} \label{eqn:pk_fit}
    P_{\rm i,fit}(k,z,\Sigma) = P(k,z) \left[1 + {\cal A}_i \cos\left(\omega_i \kappa_X + \phi_i\right){\cal D}(k,z,\Sigma)\right] \,,
\end{equation}
where $P(k,z)$ is the non-linear matter power spectrum for a $\Lambda$CDM cosmology from the 
simulations assuming that the small-scales enhancement of the matter power spectrum and the BAO 
feature smoothing due to non-linear effects is the same as in $\Lambda$CDM cosmology for this class 
of models. 
$\kappa_{\rm lin} \equiv k/k_*$ for linear oscillations \eqref{eqn:pk_lin} and 
$\kappa_{\rm log} \equiv \log \left(k/k_*\right)$ for logarithmic oscillations \eqref{eqn:pk_log}.
${\cal D}(k,z,\Sigma)$ is the damping function to model the damping of superimposed oscillations on 
the matter power spectrum due to non-linear effects. 
Analogously to the damping used for BAO \cite{Seo:2007ns}, we parameterize the damping 
function  with a Gaussian damping as:
\begin{equation} \label{eqn:damping}
    {\cal D}(k,z,\Sigma) = e^{-k^2\Sigma^2(z)/2} \,,
\end{equation}
where $\Sigma$ is the redshift-dependent parameter that we fit with our simulations.

The 6 best fitting parameters for the linear feature models given in Tab.~\ref{tab:osc_models} are:
\begin{align*}
    &\Sigma_{\rm lin} (z) = \left[12.23,  8.00,  5.70,  4.40,  3.59,  3.05\right]\ \text{Mpc} \,,\\
    &\Sigma_{\rm lin} (z) = \left[12.26,  8.02,  5.73,  4.42,  3.61,  3.06\right]\ \text{Mpc} \,,\\
    &\Sigma_{\rm lin} (z) = \left[12.20,  7.96,  5.66,  4.36,  3.55,  3.01\right]\ \text{Mpc} \,,\\
    &\Sigma_{\rm lin} (z) = \left[12.26,  7.99,  5.68,  4.38,  3.57,  3.02\right]\ \text{Mpc} \,,\\
    &\Sigma_{\rm lin} (z) = \left[12.55,  8.23,  5.88,  4.54,  3.72,  3.17\right]\ \text{Mpc} \,,\\
    &\Sigma_{\rm lin} (z) = \left[12.54,  8.21,  5.85,  4.53,  3.71,  3.16\right]\ \text{Mpc} \,,\\
\end{align*}
where different values inside the square brackets refer to different redshift, i.e. $z=0,1,2,3,4,5$.
Fitting simultaneously the 6 best-fit, we find:
\begin{equation} \label{eqn:sigma_lin}
    \Sigma_{\rm lin} (z) = \left[12.34,  8.07,  5.75,  4.44,  3.62,  3.08\right]\ \text{Mpc} \,.
\end{equation}

\begin{figure}
\centering
\includegraphics[width=.48\textwidth]{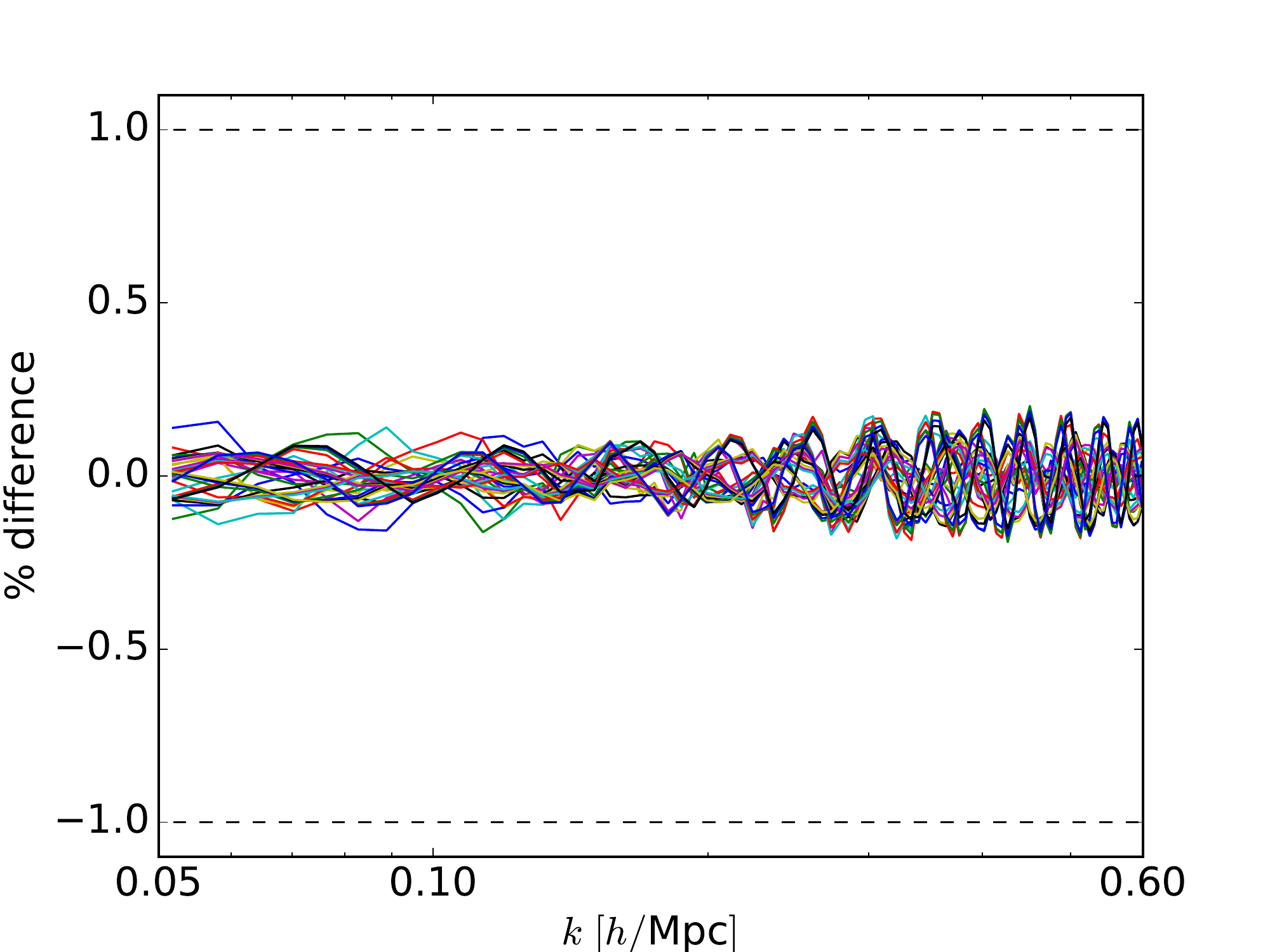}
\includegraphics[width=.48\textwidth]{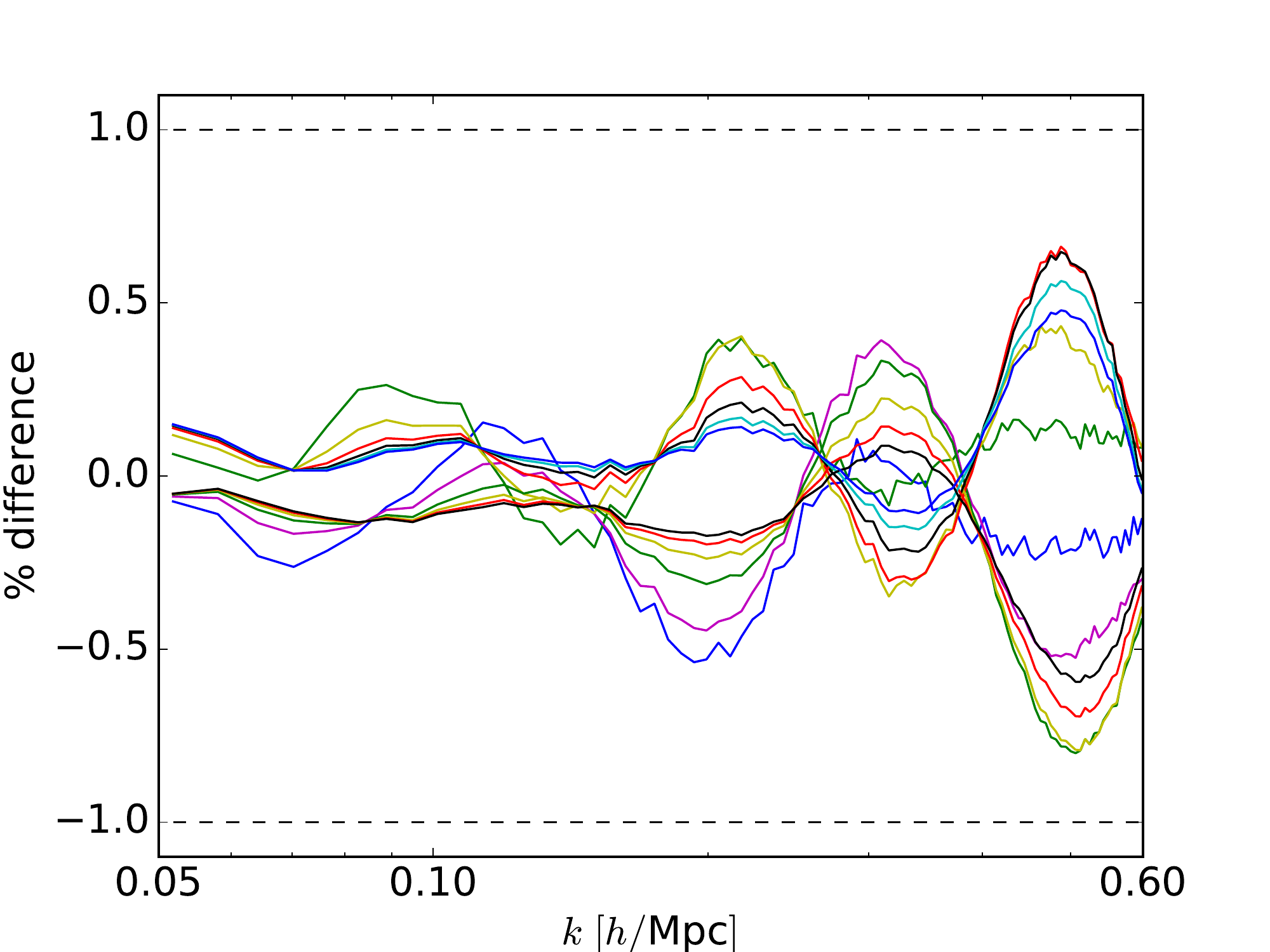}
\caption{Percentage differences between non-linear matter power spectrum reconstructed with our 
semi-analytical template \eqref{eqn:pk_fit} and the non-linear matter power spectrum obtained from 
the simulations. We show the results from the Gaussian damping function \eqref{eqn:damping} 
on the left panel for the linear model and on the right panel for the logarithmic model. 
The semi-analytical non-linear matter power spectra have been calculated 
using the simultaneous best-fit of the damping parameter $\Sigma$ \eqref{eqn:sigma_lin}-\eqref{eqn:sigma_log}.}
\label{fig:pk_diff}
\end{figure}

For the 4 best-fit of the logarithmic model \eqref{eqn:pk_log} we obtain:
\begin{align*}
    &\Sigma_{\rm log} (z) = \left[11.30,  6.68,   4.47,  3.37,  2.71,  2.27\right]\ \text{Mpc}  \,,\\
    &\Sigma_{\rm log} (z) = \left[11.23,  6.35,   4.09,  2.99,  2.34,  1.91\right]\ \text{Mpc}  \,,\\
    &\Sigma_{\rm log} (z) = \left[12.40,  7.78,   5.35,  4.06,  3.29,  2.77\right]\ \text{Mpc}  \,,\\
    &\Sigma_{\rm log} (z) = \left[13.20,  8.32,   5.74,  4.36,  3.53,  2.98\right]\ \text{Mpc}  \,,\\
\end{align*}
and fitting simultaneously the 4 best-fit, we find
\begin{equation} \label{eqn:sigma_log}
    \Sigma_{\rm log} (z) = \left[11.96,  7.26,  4.90,  3.72,  3.02,  2.55\right]\ \text{Mpc} \,.
\end{equation}
In Figs.~\ref{fig:LINfit}-\ref{fig:LOGfit}, we show the comparison between the non-linear matter 
power spectrum for $\Lambda$CDM obtained from \texttt{CAMB} with undamped superimposed oscillations 
in green, the non-linear matter power spectrum obtained from the simulations in blue, and the 
non-linear matter power spectrum for $\Lambda$CDM obtained from \texttt{CAMB} with superimposed 
oscillations obtained with our fit in dashed magenta for the linear and logarithmic models, 
respectively with the best-fit \eqref{eqn:sigma_lin} and \eqref{eqn:sigma_log}. The fit with 
the Gaussian envelope in Eq. \eqref{eqn:damping} provides an excellent fit to the simulations 
with relative differences lower than $0.2\%$ for the linear model and $0.6\%$ for the logarithmic 
one, up to $k \leq 0.6\ h$/Mpc, see left panels on Fig.~\ref{fig:pk_diff}.
Note that the absolute variance on the best-fit estimated for $\Sigma_{\rm lin}$ 
and $\Sigma_{\rm log}$ is smaller then 0.02.

\begin{figure}
\centering
\includegraphics[width=.48\textwidth]{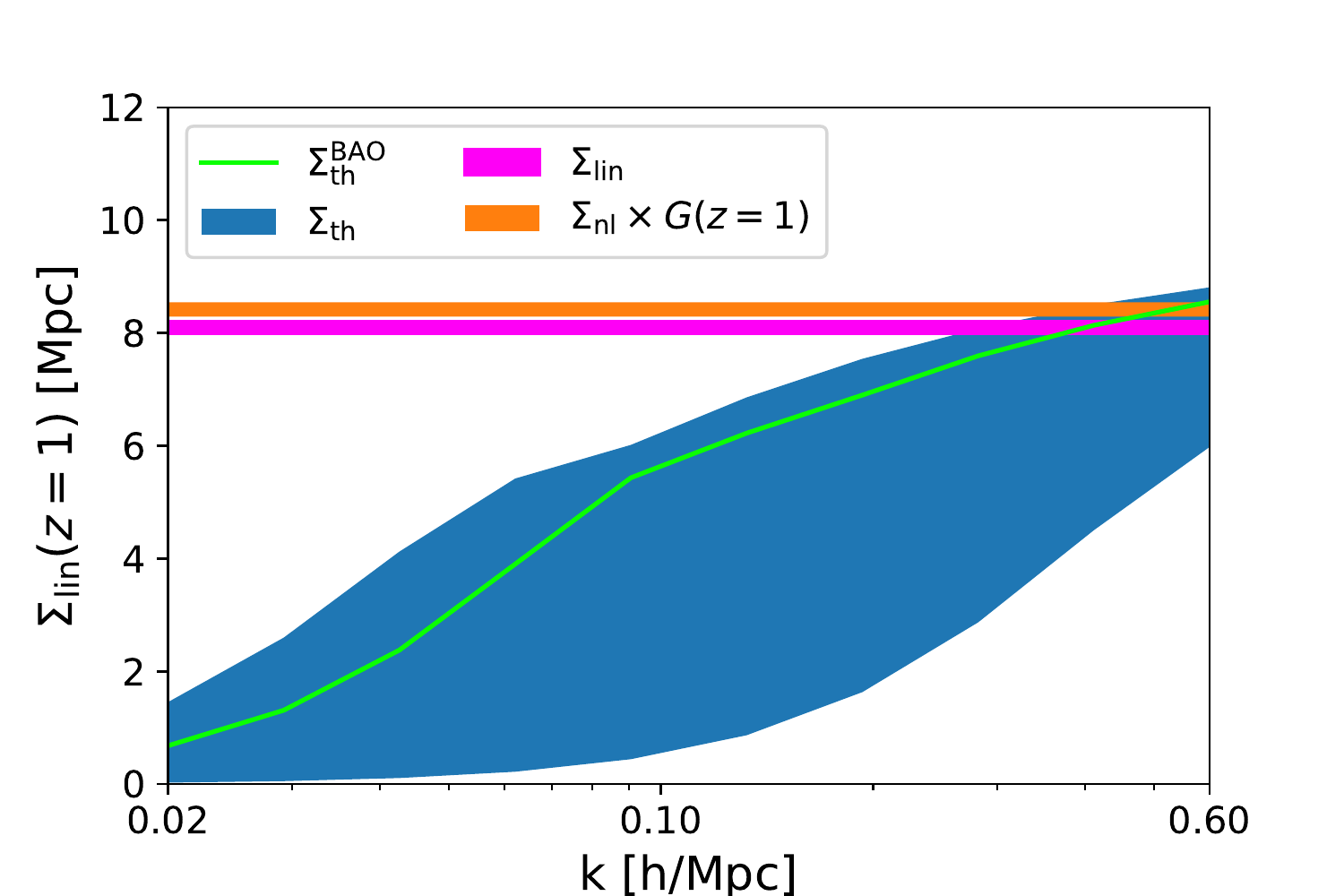}
\includegraphics[width=.48\textwidth]{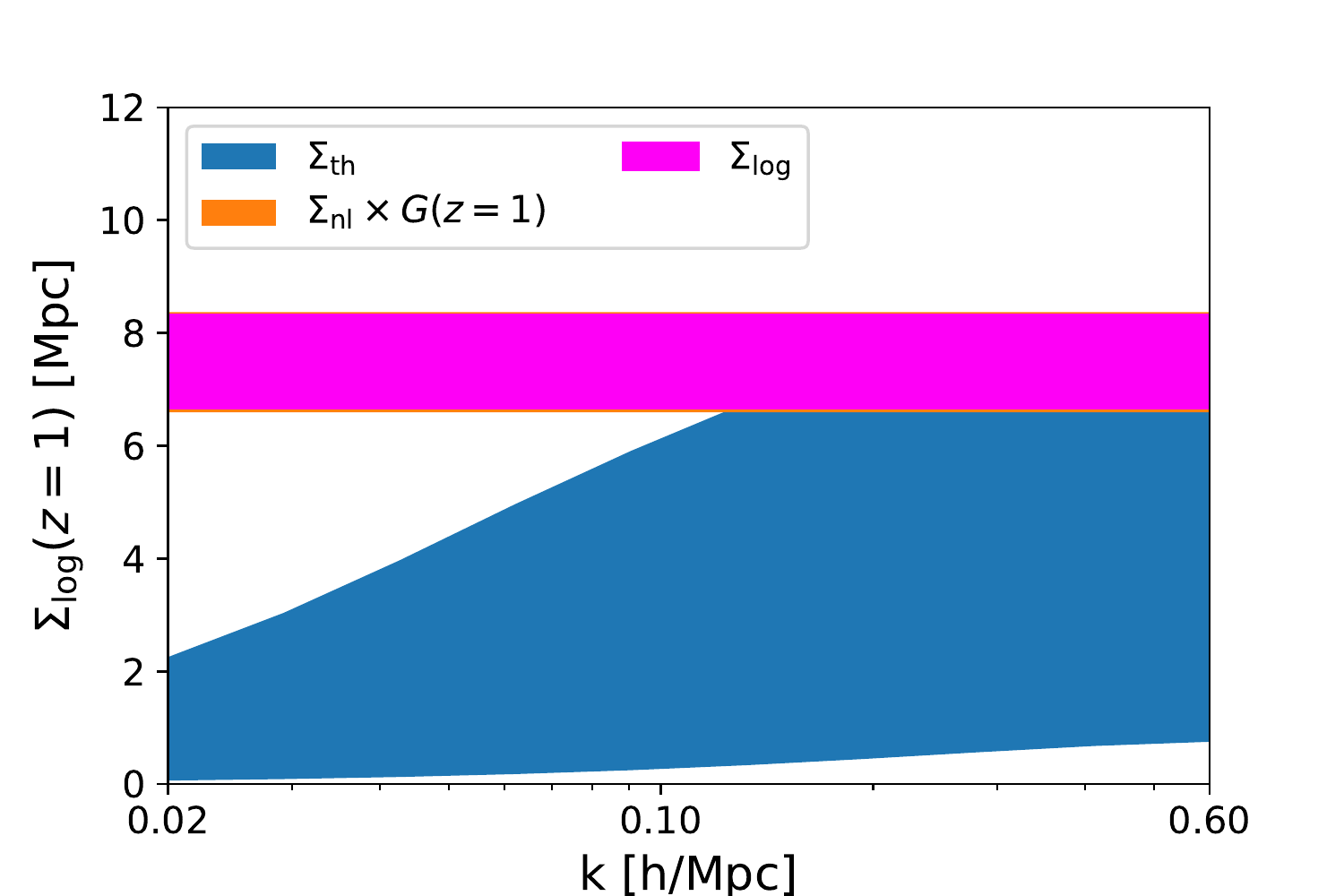}
\caption{Comparison of our fit for the non-linear damping parameter $\Sigma$ for linear (left panel) and logarithmic 
(right panel) template of primordial oscillations at redshift $z=1$ with theoretical predictions to leading order 
\cite{Beutler:2019ojk,Vasudevan:2019ewf}. 
We display $\Sigma (z=1)$ in magenta and $\Sigma_{\rm nl} G(z)$ in orange where 
$\Sigma (z=1)$ and $\Sigma_{\rm nl} \equiv \Sigma (z=0)$ have been fitted to the N-body simulations. 
The blue band shows the theoretical prediction to leading order \cite{Beutler:2019ojk,Vasudevan:2019ewf} 
when the separation scale $\Lambda$ varied in the range $(0.1-0.7)\,k$ for the same frequencies of 
our simulations. 
We include also the theoretical prediction for the BAO damping with $\omega \sim 110\ \text{Mpc}/h$ 
(green line) and $\Lambda = 0.5\, k$ according to \cite{Baldauf:2015xfa}.} \label{fig:Sigma}
\end{figure}

We then want to compare our findings with the analytic results previously obtained in \cite{Beutler:2019ojk,Vasudevan:2019ewf}. The redshift behaviour from our simulations is 
very well reproduced by the growth factor $G(z)$, i.e. $\Sigma(z) = \Sigma_{\rm nl} G(z)$, as 
analytically studied in \cite{Seo:2007ns,Blas:2016sfa}.
Based on our Eqs.~\eqref{eqn:pk_fit}-\eqref{eqn:damping}, we compare our results for 
$\Sigma(z)$ with the leading order from perturbation theory \cite{Beutler:2019ojk,Vasudevan:2019ewf}:
\begin{equation}
    \Sigma_{\rm th}^2(k,z) = \frac{1}{3\pi^2} \int_0^\Lambda \dd q \left[ 1-j_0(q\omega)+2 j_2(q\omega)\right]
    P_{\rm lin}(q,z) \,,
\end{equation}
where the separation scale $\Lambda$ is suggested to be scale dependent $\Lambda = \epsilon k$ 
with $\epsilon \in [0.1,0.7]$ \cite{Baldauf:2015xfa,Vasudevan:2019ewf} and $j_n$ are the spherical 
Bessel function. 
For the linear template we have $\omega \to \omega_{\rm lin}/(0.05\ \text{Mpc}^{-1})$ and for the logarithmic template 
$\omega \to \omega_{\rm log}/k$, as derived in \cite{Beutler:2019ojk,Vasudevan:2019ewf}. 
Fig.~\ref{fig:Sigma} shows 
how our estimate for $\Sigma(z)$ is consistent with the analytic estimates to 
leading order for the linear and logarithmic wiggles according to \cite{Beutler:2019ojk,Vasudevan:2019ewf}.

\section{Comparison with the BAO signal} \label{sec:BAO}

We now want to compare the linear template with the BAO signal.
The matter power spectrum can be modeled by a smooth power spectrum without wiggles (nw) plus the 
BAO spectrum like:
\begin{equation}
	P(k,z) \approx P_{\rm nw}(k,z) \left[1 + A_{\rm BAO}(k)\sin\left(k r_s(z) + \phi\right) \right] \,.
\end{equation}
The BAO signal in Fourier space looks very similar to the oscillatory pattern induced on the 
matter power spectrum by the primordial linear oscillations \eqref{eqn:pk_lin} with a frequency 
 $\log_{10} \left(\omega_{\rm lin}\right) \sim 0.87$.

\begin{figure}[t]
\centering
\includegraphics[width=1.1\textwidth]{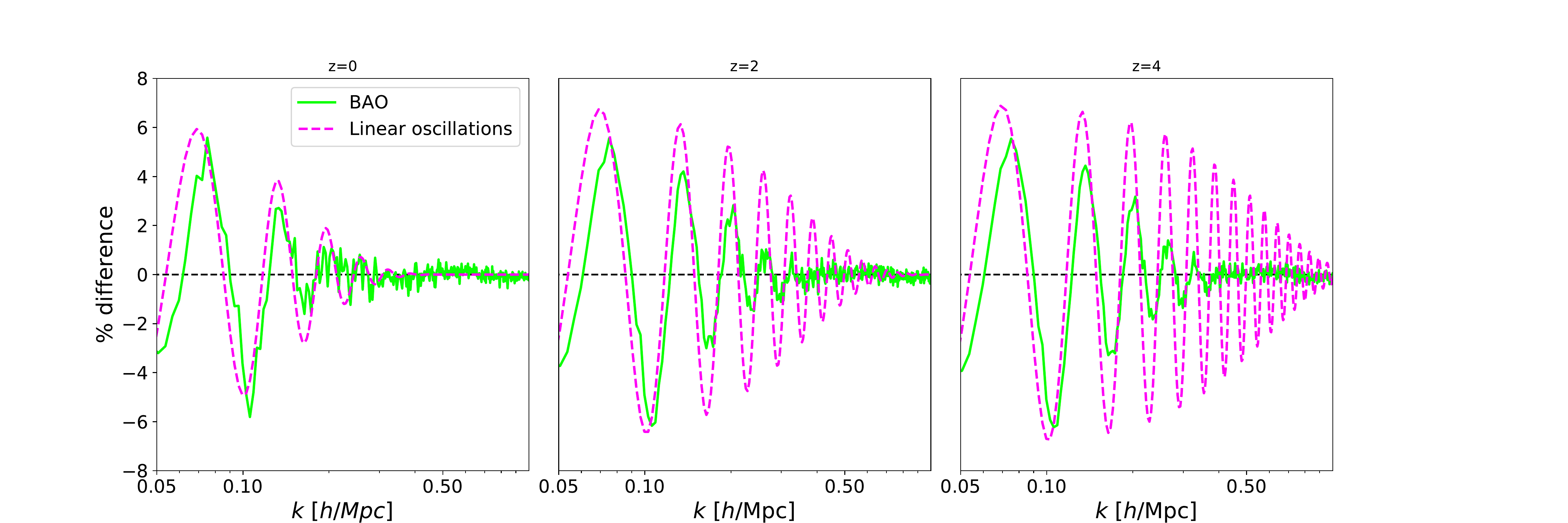}
\caption{Comparison between the BAO signal at $z=0,\,2,\,4$ (green solid line) 
extracted from our $\Lambda$CDM simulation and our fit for the linear template with ${\cal A}_{\rm lin}=0.07$, $\log_{10} \left(\omega_{\rm lin}\right) = 0.87$, 
$\phi/(2\pi)=0.4$ (magenta dashed line). 
The green curve shows the relative differences between the non-linear matter power spectrum 
with and without BAO wiggles for $\Lambda$CDM (both without superimposed oscillations).
The dashed magenta curve shows the relative differences between the non-linear matter power 
spectrum with and without superimposed linear oscillations \eqref{eqn:pk_lin} (both with BAO). 
We perform the BAO signal subtraction with a polynomial method following Ref.~\cite{Hinton:2016atz}. 
We tune it in order to have a good BAO signal subtraction and non-distortion of the broadband power spectrum.
}
\label{fig:BAOcomparison2}
\end{figure}

As can be seen in Fig.~\ref{fig:BAOcomparison2}, even with a fine tuned frequency, the linear 
template is different from the BAO signal at early times, i.e. $z=4$: we can see the footprints of the Silk damping on the BAO signal \cite{Silk:1967kq}, 
but not on the primordial oscillations.

\begin{figure}[t]
\centering
\includegraphics[width=1.1\textwidth]{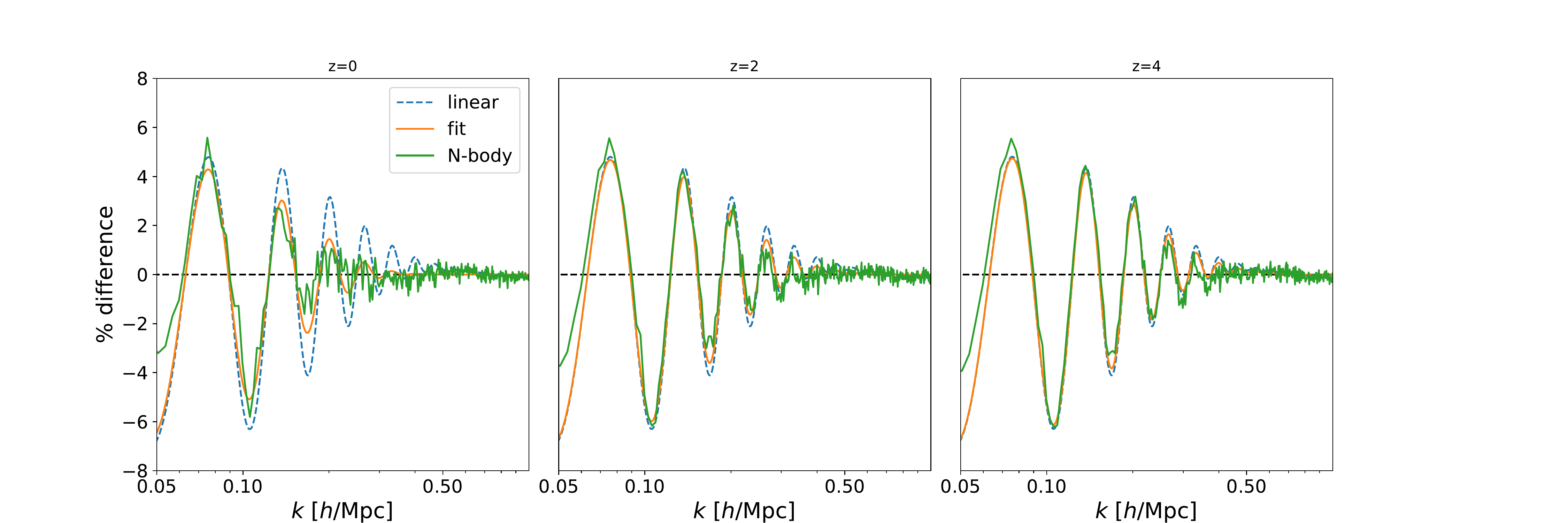}
\caption{Relative differences between the non-linear matter power spectrum for 
$\Lambda$CDM with the smooth power spectrum without BAO. The blue dashed curve shows the relative 
differences for the linear matter power spectrum, the orange curve shows the relative differences 
for the non-linear matter power spectrum obtained with the Gaussian damping \eqref{eqn:damping} 
with the damping parameter $\Sigma_{\rm BAO}$ \eqref{eqn:sigma_bao} fitted to the N-body simulations, 
and the green curve shows the relative differences for the non-linear matter power spectrum extracted 
from the N-body simulations.}
\label{fig:BAOdamping}
\end{figure}

As consistency check, we extract the damping of the BAO from our $\Lambda$CDM N-body 
simulations. We fit BAO non-linear damping by using the Gaussian envelope \eqref{eqn:damping}:
\begin{equation} \label{eqn:sigma_bao}
    \Sigma_{\rm BAO} (z) = \left[13.90,  9.15,  6.82,  5.66,  5.01,  4.62\right]\ \text{Mpc} \,.
\end{equation}
with an absolute variance on the $\Sigma_{\rm BAO}$ estimated smaller than 0.7, which is consistent 
with results in literature (see for instance Ref.~\cite{Ding:2017gad}).
See Fig.~\ref{fig:BAOdamping} for a comparison of the linear and non-linear BAO signal.

\section{Forecast for future galaxy surveys} \label{sec:analysis}

We describe in this section the Fisher matrix methodology for the galaxy clustering (GC) and 
the CMB used for our forecasts. We also describe the specifications for the different experiments 
considered: Euclid-like, Subaru Prime Focus Spectrograph (PFS), {\em Planck}-like, and a 
CMB cosmic-variance (CV) experiment.

\subsection{Galaxy power spectrum}

We measure galaxy positions in angular and redshift coordinates and not the position 
in comoving coordinates, i.e. the true galaxy power spectrum is not a direct observable.
We use a model for the observed galaxy power spectrum based on \cite{Seo:2003pu,Song:2008qt,Wang:2012bx}:
\begin{equation} \label{eqn:Pobs}
    P_{\rm obs}(k^{\rm ref}_\perp, k^{\rm ref}_\parallel,z) = 
    \left[ \frac{D^{\rm ref}_{\rm A}(z)}{D_{\rm A}(z)} \right]^2
    \frac{H(z)}{H^{\rm ref}(z)} 
    F_{\rm FoG}(k,z)
    \frac{P_{\rm fit}(k,z)}{\sigma_8^2(z)} + P_{\rm shot}(z) \,,
\end{equation}
where $D_\mathrm{A} = r(z)/(1+z)$ is the angular diameter distance, $r(z)$ is the comoving distance, 
$H(z) = \dot{a}/a$ is the Hubble parameter, $k^2=k_\perp^2+k_\parallel^2$, and 
$\mu=k_\parallel/k=\hat{\bf r}\cdot\hat{\bf k}$. This is connected 
to the true galaxy power spectrum via a coordinate transformation \cite{Alcock:1979mp}:
\begin{equation}
    k^{\rm ref}_\perp = \frac{D_{\rm A}(z)}{D^{\rm ref}_{\rm A}(z)}\,k_\perp\,,
    \qquad k^{\rm ref}_\parallel=\frac{H^{\rm ref}(z)}{H(z)}\,k_\parallel\,.
\end{equation}
In Eq.~\eqref{eqn:Pobs}, $P_{\rm shot}$ is the shot noise and we model the redshift-space distortions (RSD) 
as:
\begin{equation}
    F_{\rm FoG}(k,z) = \frac{\left[b(z)\sigma_8(z)+f(z)\sigma_8(z)\mu^2\right]^2}{1+k^2\mu^2\sigma^2_{r,p}/2}
    e^{-k^2\mu^2\sigma_{r,z}^2} \,,
\end{equation}
where $b(z)$ is the linear clustering bias, $f(k,z)= \dd\ln {G}(k,z) / \dd\ln a$ is the growth rate, and 
$G(k,z)$ is the growth factor.
Here the numerator is the linear RSD \cite{Kaiser:1987qv,Hamilton:1997zq}, which takes into account 
the enhancement due to large-scale peculiar velocities. The Lorentzian denominator models the non-linear 
damping due to small-scale peculiar velocities, where $\sigma_{r,p}$ is the distance dispersion:
\begin{equation}
{\sigma_{r,p}(z)} = \frac{\sigma_{p}(z)}{H(z)a(z)} \,, 
\end{equation}
corresponding to the physical velocity dispersion $\sigma_p$. We choose a value of 
$\sigma_p = 290$\,km/s as our fiducial \cite{Wang:2012bx}.
An additional exponential damping factor is added to account for {the error $\sigma_z$} in the 
determination of the redshift of sources, where:
\begin{equation}
\sigma_{r,z}(z) = \frac{\partial r}{\partial z} \sigma_z = \frac{c}{H(z)}\sigma_z \,.
\end{equation}

We model the smearing of the BAO feature according to \cite{Seo:2007ns,Wang:2012bx}:
\begin{align}
P_\mathrm{dw}(k,\mu,z) = &P_\mathrm{m}(k,\mu,z)e^{-\Sigma_{\rm BAO}^2(z)k^2/2} \notag\\
&+ P_\mathrm{nw}(k,\mu,z)\left(1-e^{-\Sigma_{\rm BAO}^2(z)k^2/2}\right) ,
\end{align}
where $\Sigma_{\rm BAO}(z) \equiv \Sigma_{\rm BAO}G(z)$ with $\Sigma_{\rm BAO}(z=0)=9.3 \text{ Mpc}/h$.
Here $P_\mathrm{dw}$ is dressed with the damped primordial oscillation 
fitted to the N-body simulations according to Eq.~\eqref{eqn:pk_fit}.

Finally, the finite size of a galaxy survey and the survey window function introduce couplings between 
different modes $k$ and, as a consequence, discrete bandpowers should be considered in the analysis in 
order to avoid these correlations. 
We model the observed matter power spectrum \eqref{eqn:Pobs} in bandpowers averaged over a bandwidth 
$\Delta k$ with a top-hat window function as in Refs.~\cite{Huang:2012mr,Beutler:2019ojk}:
\begin{equation} \label{eqn:tophat}
    \hat{P}_{\rm obs}(k_i,z) = \frac{1}{\Delta k}\int_{k_i-\Delta k/2}^{k_i+\Delta k/2} \dd k' P_{\rm obs}(k',z) \,.
\end{equation}

\subsection{Fisher analysis}

We follow the same approach as in Ref.~\cite{Ballardini:2017qwq} (see also 
Refs.~\cite{Tegmark:1997rp,Seo:2003pu}). The Fisher matrix for the observed matter power 
spectrum \eqref{eqn:Pobs}, for a $i$-th redshift bin, is given by:
\begin{equation} \label{eqn:fisher}
    F_{\alpha\beta}^{\rm GC}(z_i) = 
    \int_{-1}^1 \dd \mu \int_{k_{\rm min}}^{k_{\rm max}} \frac{k^2\ \dd k}{8\pi^2}
    \frac{\partial \ln P_{\rm obs}(k,\mu,z_i)}{\partial \alpha}
    \frac{\partial \ln P_{\rm obs}(k,\mu,z_i)}{\partial \beta}
    V_{\rm eff}(z_i) \,,
\end{equation}
where $k$ and $\mu$ are the ones related to the reference cosmology, 
$\partial P_{\rm obs}/\partial \alpha$ is the derivative with respect to the $\alpha$ element 
in the cosmological parameter vector {\boldmath$\theta$}.
The effective volume in the $i$-th redshift bin, is given by \cite{Feldman:1993ky}:
\begin{equation} \label{eqn:Veff}
    V_{\rm eff}(k,\mu,z_i) \simeq V_{\rm surv}(z_i) 
    \left[\frac{n(z_i)P_{\rm obs}(k,\mu;z_i)}{n(z_i)P_{\rm obs}(k,\mu,z_i) + 1}\right]^2 \,,
\end{equation}
where $V_{\rm surv}(z_i)$ is the comoving volume in the $i$-th redshift bin.

The full set of parameters {\boldmath$\theta$} includes the standard shape parameters 
$\big\{\omega_c,\ \omega_b,\ h,\ n_s\big\}$, the redshift-depedent parameters
$\big\{H, D_\mathrm{A}, \log\left(f\sigma_8\right)\big\}_{z_i}$, the redshift-depedent 
nuisance parameters $\big\{\log\left(b\sigma_8\right),\ P_{\rm shot},\ \sigma_p\big\}_{z_i}$
together with the three extra parameters of the primordial oscillatory feature model 
$\big\{{\cal A}_X, \log_{10} (\omega_X), \phi/(2\pi)\big\}$ (see Sec.~\ref{sec:models}). 
After marginalizing over the nuisance parameters, we project the redshift-dependent parameters 
on the final set of cosmological parameters
\begin{equation}
    \big\{\omega_c,\ \omega_b,\ h,\ n_s,\ \ln\left(10^{10}A_s\right),{\cal A}_X, \log_{10} (\omega_X), \phi/(2\pi)\big\} \,.
\end{equation}

The Fisher matrix for CMB angular power spectra (temperature and E-mode polarization) 
is~\cite{Knox:1995dq,Jungman:1995bz,Seljak:1996ti,Zaldarriaga:1996xe,Kamionkowski:1996ks}:
\begin{equation}
    F_{\alpha\beta}^{\rm CMB} = f_{\rm sky}\sum_\ell 
    \frac{2\ell+1}{2}\text{tr}
    \left[{\bf C}_{\ell,\alpha}{\bf \Sigma}_{\ell}{\bf C}_{\ell,\beta}{\bf \Sigma}_{\ell}\right] \,,
\end{equation}
where ${\bf C}_{\ell}$ is the covariance matrix, 
${\bf C}_{\ell,\alpha} \equiv \partial {\bf C}_{\ell}/\partial \alpha$ is the derivative 
with respect to the $\alpha$ element in the cosmological parameter vector {\boldmath$\theta$}, and 
${\bf \Sigma}_{\ell} \equiv \left({\bf C}_{\ell}+{\bf N}_{\ell}\right)^{-1}$ is the inverse of 
the total noise matrix with ${\bf N}_{\ell}$ the diagonal noise matrix.
The effective noise $N^{X}_{\ell}$ is the instrumental noise convolved with the beams of different 
frequency channels \cite{Ballardini:2016hpi}.
We adopt the specifications denoted as CMB-1 in \cite{Ballardini:2016hpi} for a {\em Planck}-like 
sensitivity, which reproduce uncertainties for standard cosmological parameters similar to those which 
can be obtained by {\em Planck} \cite{Akrami:2018vks}.

We study the predictions for a CV-CMB experiment considering the specifications of 
$f_{\rm sky}= 0.7$, and a multipole range from $\ell_{\rm min} = 2$ up to $\ell_{\rm max} = 2500$.

The full set of parameters {\boldmath$\theta$} for the CMB includes 
\begin{equation}
    \big\{\omega_c,\ \omega_b,\ h,\ n_s,\ \tau,\ \ln\left(10^{10}A_s\right),{\cal A}_X, \log_{10} (\omega_X), \phi_X/(2\pi)\big\} \,.
\end{equation}
We marginalize over $\tau$ the Fisher matrix of the CMB before combining it with the one of the GC.

\subsection{Galaxy clustering specifications} \label{sec:specs}

We focus on two spectroscopic galaxy surveys.
First, we consider a Euclid-like spectroscopic survey that will probe $f_{\rm sky} = 15,000$ deg$^2$ 
over a redshift range $0.9 \leq z \leq 1.8$ divided in 9 tomographic redshift bins  
equally spaced.
We adopt the predicted redshift distribution of the number counts of H$\alpha$-emitting galaxies, 
dN/dz, per square degree for Euclid-like H$\alpha$-selected survey from Ref.~\cite{Merson:2017efv} 
with H$\alpha$ + [NII] blended flux limits of $2 \times 10^{-16}$ erg s$^{-1}$ cm$^{-2}$ and dust 
method from \cite{Calzetti:1999pg}, and the linear clustering bias from Ref.~\cite{Merson:2019vfr}.
Secondly, we consider the Subaru Prime Focus Spectrograph (PFS) which will map emission line galaxies 
spanning a redshift range $0.8 < z < 2.4$ over 1,464 deg$^2$ \cite{Ellis:2012rn}.
In this case, we assume a redshift accuracy of $\sigma_z = 0.001$ for both the two experiments.

\section{Results} \label{sec:results}

\begin{figure}
\centering
\includegraphics[width=.48\textwidth]{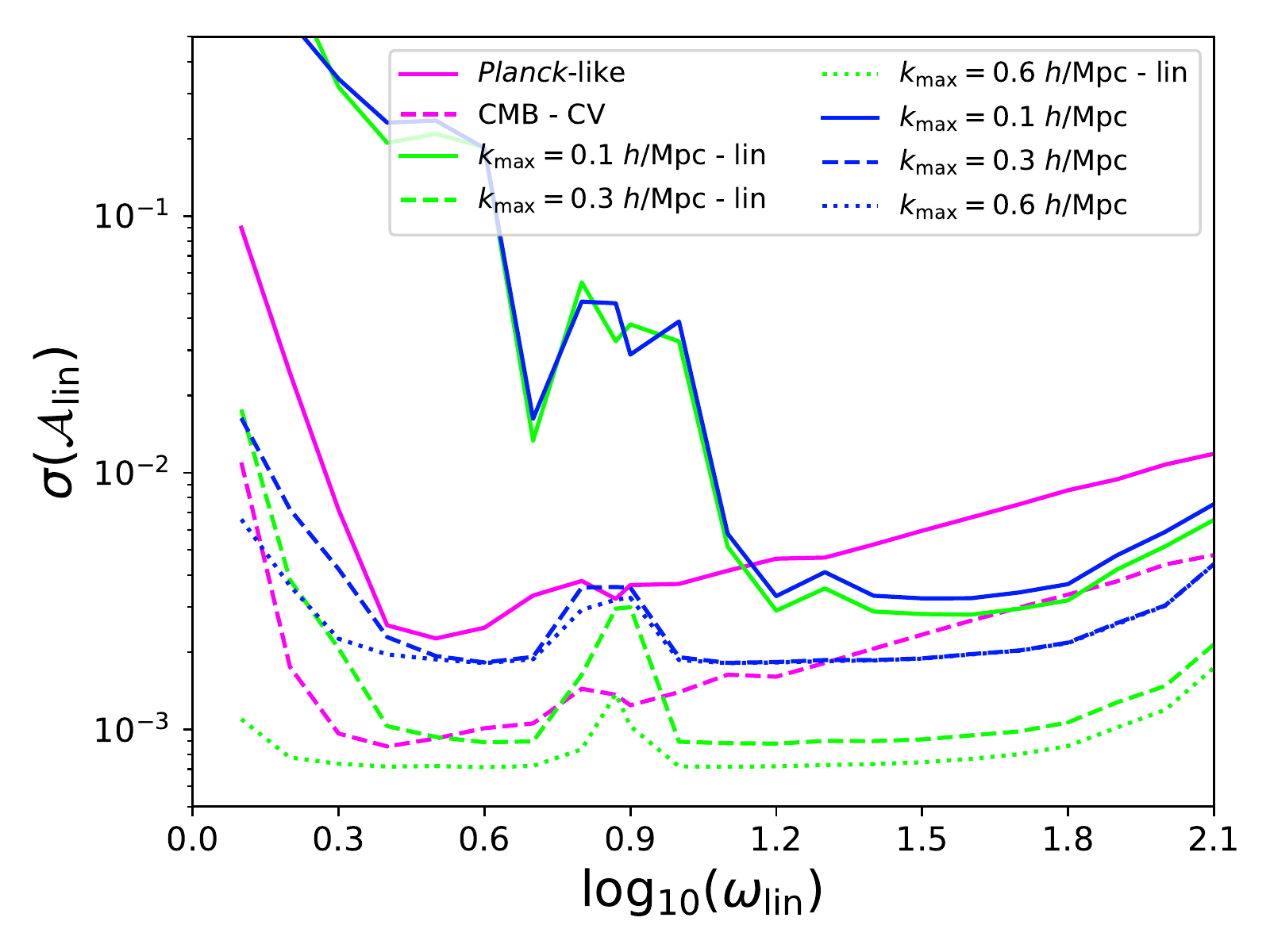}
\includegraphics[width=.48\textwidth]{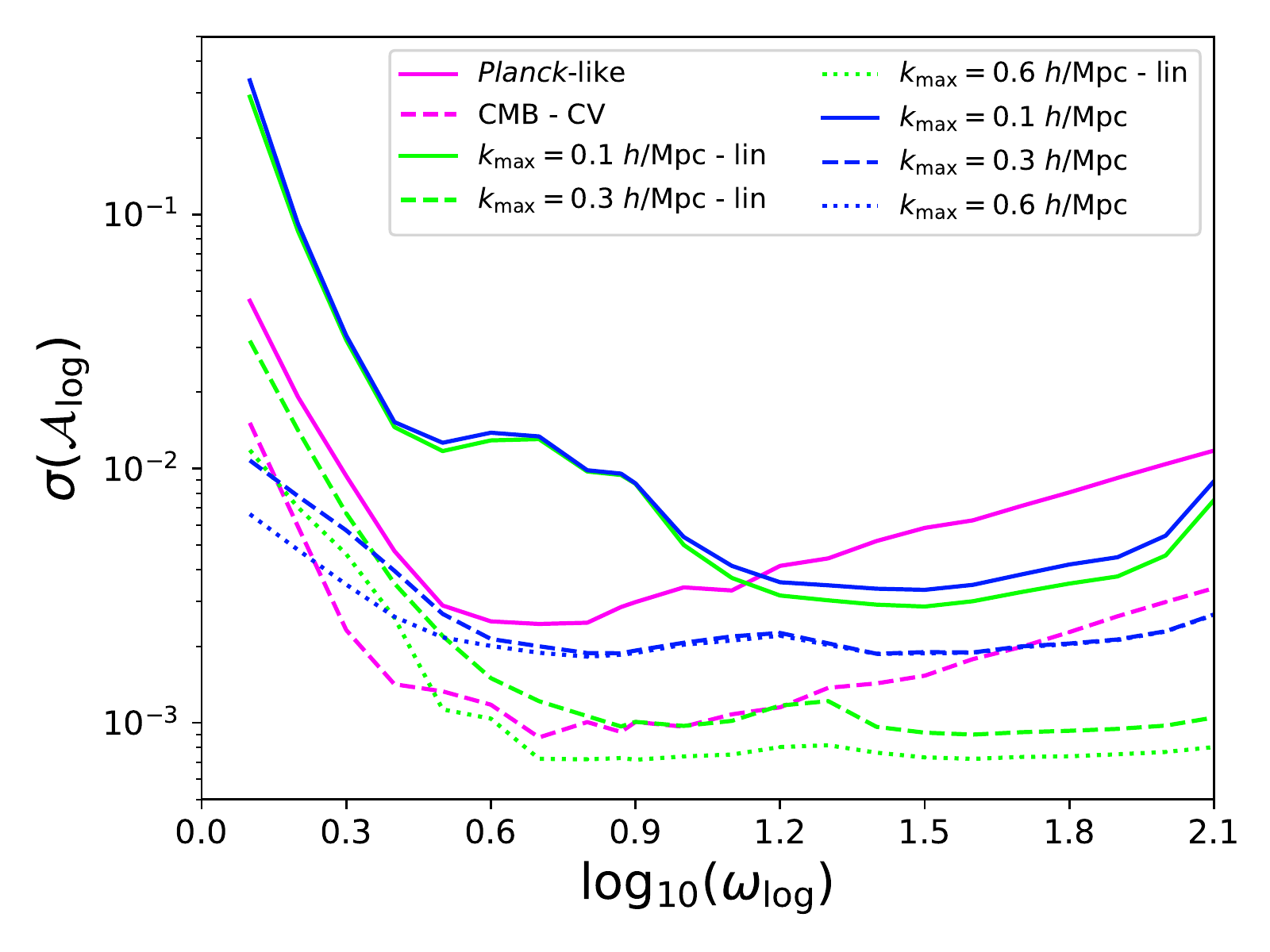}
\includegraphics[width=.48\textwidth]{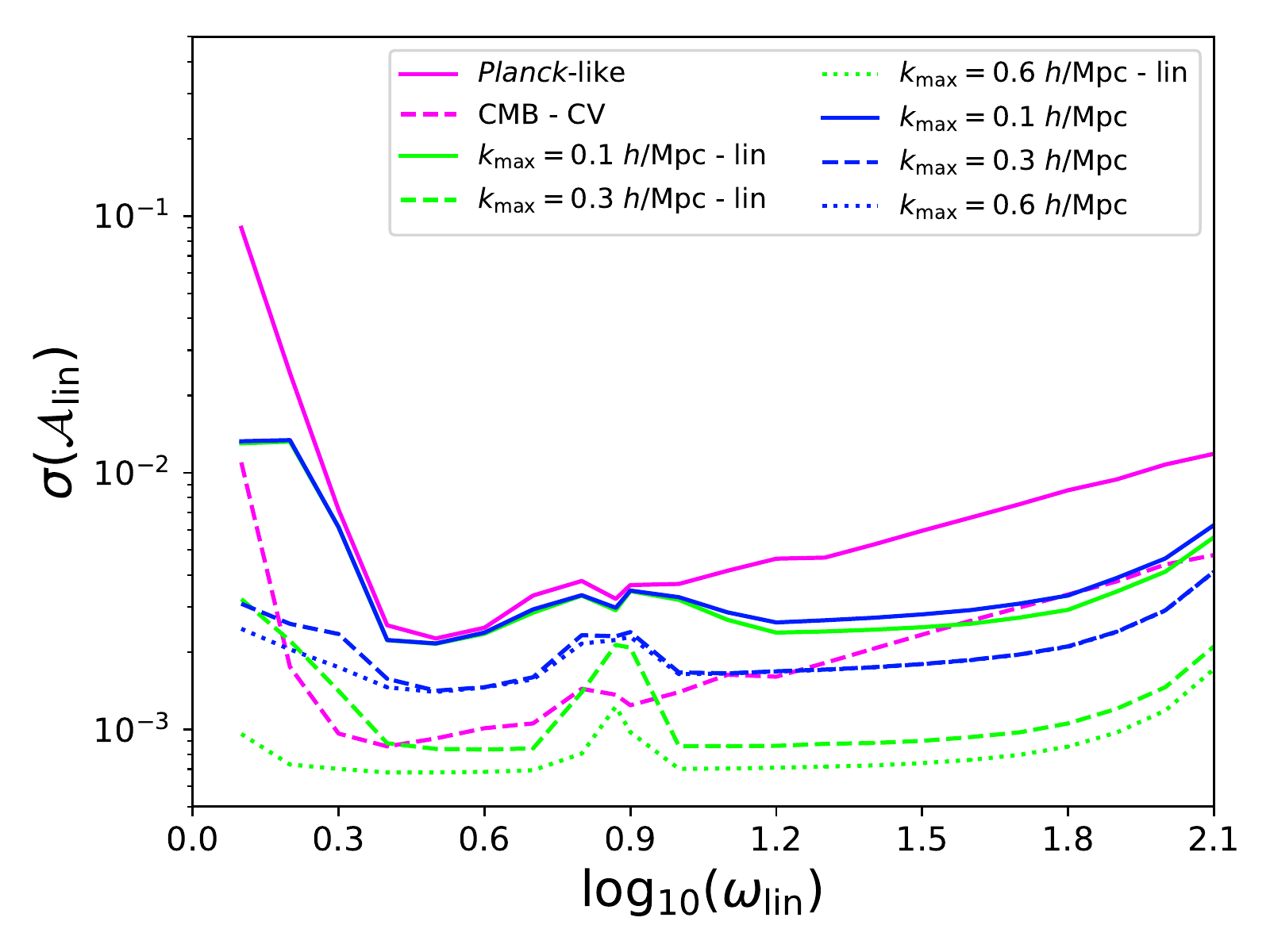}
\includegraphics[width=.48\textwidth]{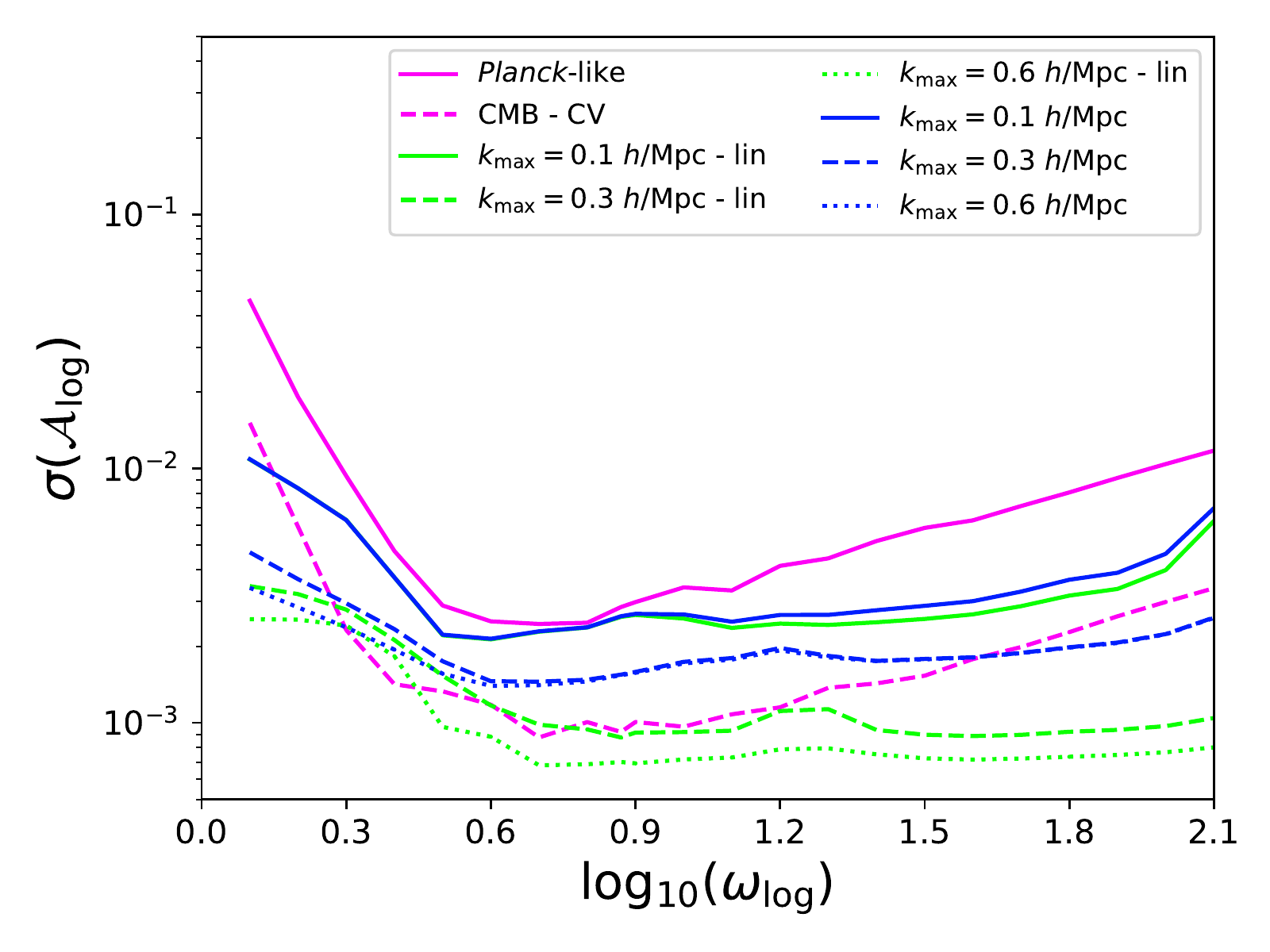}
\caption{Marginalized uncertainties on ${\cal A}_X$ as function of the frequency $\omega_X$ 
for the linear model (left panel) and the logarithmic model (right model). 
We show uncertainties for CMB-1 (magenta) and for Euclid-like (combined with {\em Planck}-like information) 
in the top (bottom) panels for different maximum wavenumber considered $k_{\rm max}=0.1,0.3,0.6\ h$/Mpc 
(solid, dashed, dotted).
Green lines refer to the linear matter power spectrum in \eqref{eqn:Pobs} and 
the blue lines to the non-linear matter power spectrum. CMB CV uncertainties are in dashed-magenta.}
\label{fig:sigmaA}
\end{figure}

We now discuss our results for the two oscillatory models considered (see Sec.~\ref{sec:models} 
for the parameterizations).
The marginalized 68\% constraints on the amplitude ${\cal A}$ 
(for different values of $\log_{10}\left(\omega_X\right)$ 
around the best-fit ${\cal A}_X = 0.03$ and $\phi_X/(2\pi) = 0.2$)
for a Euclid-like experiment 
are shown in Fig.~\ref{fig:sigmaA} and can be summarized as follows:
\begin{itemize}
    \item using the linear matter power spectrum in \eqref{eqn:Pobs}, we recover uncertainties 
    on the amplitude consistent with the ones obtained in Ref.~\cite{Chen:2016vvw} for the linear 
    model \eqref{eqn:pk_lin} and in Refs.~\cite{Chen:2016vvw,Ballardini:2016hpi} for the logarithmic 
    model \eqref{eqn:pk_log} for $k_{\rm max}=0.1\ h$/Mpc.
    We find similar uncertainties using the non-linear matter power spectrum when $k_{\rm max}=0.1\ h$/Mpc. This confirms the validity of using the linear theory when restricting to these scales;
    \item when $k_{\rm max}=0.6\ h$/Mpc, we find that Euclid-like can decrease the uncertainties 
    $\sigma\left({\cal A}\right)$ by a factor 2 when $\log_{10}\left(\omega_X\right) > 1.0$ thanks 
    to our modelling of non-linear effects. Note that incorrectly by discarting non-linear 
    effects one would get much tighter constraints.
    We find that a Euclid-like survey can lead to uncertainties which improve on a CV-CMB for 
    $\log_{10}\left(\omega_{\rm lin}\right) > 1.3$ and $\log_{10}\left(\omega_{\rm log}\right) > 1.6$;
    \item another interesting aspect is that the improvement in term of uncertainties saturates 
    at $k_{\rm max}=0.3\ h$/Mpc for the model considered. This trend is due to the 
    non-linear damping which smooths most of the oscillations for $k > 0.3\ h$/Mpc for $z < 2$;
    \item finally we see that for the linear model the uncertainties for frequencies 
    around the BAO frequency $\log_{10}\left(\omega_{\rm lin}\right) \simeq 0.87$ are sensitively 
    degraded, as pointed out in Ref.~\cite{Beutler:2019ojk}.
\end{itemize}
Our best forecasted constraint from Euclid-like GC in combining with CMB-1 ({\em Planck}-like) 
information for both the two models ($X$ = lin, log) corresponds to 
$\sigma\left({\cal A}_X\right) \simeq 0.0018$ for $\log_{10}\left(\omega_X\right) = 1.1$ 
and $\sigma\left({\cal A}_X\right) \simeq 0.0026$ for $\log_{10}\left(\omega_X\right) = 2.1$, respectively.

We then consider PFS. We find also in this case that the improvement in terms of uncertainties 
saturates around $k_{\rm max}=0.3\ h$/Mpc, even if PFS covers redshifts larger than $z=2$.
Despite the smaller sky coverage compared to Euclid (by $\sim 10\%$), we find 
similar uncertainties for PFS when combined with CMB information, 2 times larger than the uncertainties 
obtained for the same frequencies for the Euclid-like specifications.

Finally, we combine the two GC clustering experiments with a future full-sky %($f_{\rm sky} = 0.7$) 
CV-CMB experiment inspired by proposed CMB satellites 
\cite{Finelli:2016cyd,Hanany:2019lle,Delabrouille:2019thj}. Despite the large improvement by 
a factor of 3 in terms of uncertainties when we consider CMB alone (see Fig.~\ref{fig:sigmaA}), 
once we combine CMB with GC information the improvement from {\em Planck}-like to CV is minor and 
$\lesssim 10\%$ for $\log_{10}\left(\omega_X\right) > 1$.

In Fig.~\ref{fig:sigmaA}, we can see that the uncertainties for high frequencies become large.
This is due to the window function \eqref{eqn:tophat} which progressively damp 
frequencies $\omega_{\rm lin} \gtrsim 0.05/\Delta k$.
For $\Delta k \simeq 0.005\, h$/Mpc, close to the fundamental mode of BOSS and PFS, frequencies 
higher than $\omega_{\rm lin} \sim 10$ start to be damped. For the spectroscopic survey expected 
for Euclid a smaller bandwidth of $\Delta k \simeq 0.0025$ should guarantee optimal constraints 
up to $\omega_{\rm lin} \sim 20$.
For the logarithmic model the oscillations persist on small scales up to higher frequencies. 
The addition of the density field reconstruction to the analysis as done in 
Ref.~\cite{Beutler:2019ojk} can further improve the constraints.

\section{Conclusions} \label{sec:conclusion}

Global features in the primordial power spectrum 
provide a variety of information on the physics of the early Universe ranging 
from the detection of new heaviest particles, of the presence of a fast-roll stage, to fine details in 
the inflationary dynamics. They can also be used to discriminate between inflation and alternative 
scenarios in presence of signals which are oscillatory in time.

LSS experiments (also in the perspective of the next coming surveys) give the opportunity to further 
investigate the presence of any salient features in the matter power spectrum, complementing the 
constraints based on CMB anisotropy measurements to smaller scales.
In Refs.~\cite{Chen:2016vvw,Ballardini:2016hpi,Chen:2016zuu,Xu:2016kwz,Fard:2017oex,Palma:2017wxu,Beutler:2019ojk},  
it has been already pointed out the complementarity between the matter power spectrum from future 
galaxy surveys and the angular power spectrum from the measurements of CMB anisotropies in temperature 
and polarization to help in characterizing primordial features in the primordial power spectrum.
In particular, in Refs.~\cite{Chen:2016vvw,Ballardini:2016hpi} it has been shown how future 
LSS surveys will be able to improve current constraints on these oscillatory-features models just by using linear scales, i.e. $k \lesssim 0.1\ h$ Mpc$^{-1}$.

In order to study the imprints of primordial features on all scales probed by galaxy surveys, we have run a set of high-resolution DM-only cosmological simulations corresponding to different models with linear 
and logarithmic superimposed oscillations with 1,024$^3$ DM particles in a comoving box with side 
length of 1,024 Mpc$/h$ and 2,048 Mpc$/h$ (see \cite{Vlah:2015zda,LHuillier:2017lgm} for previous applications 
of N-body simulations to different models of primordial features).
Our study is important to understand the fully non-linear regime for the clustering in models with 
primordial features. Our results complement analytic approximations based on a perturbative treatment, 
see Refs.~\cite{Vlah:2015zda,Beutler:2019ojk,Vasudevan:2019ewf}, and show a compatible 
non-linear damping with respect to analytic results to leading order. We stress that these 
effects are relevant for current galaxy surveys like BOSS and eBOSS \cite{Alam:2016hwk}, DESI \cite{desi}, 
DES \cite{Elvin-Poole:2017xsf}, as well as for future experiments such as Euclid and PSF-Subaru.

After calibrating the damping of the primordial oscillations with a semi-analytical template 
\eqref{eqn:damping} against the matter power spectrum extracted from the N-body simulations 
at different redshifts, we have studied the forecasted uncertainties extending our previous 
analysis \cite{Ballardini:2016hpi} on the capability of GC up to quasi-linear scales 
$k \lesssim 0.1\ h/$Mpc to improve the uncertainties for such class of primordial models. 
The uncertainties on the amplitude of the linear (logarithmic) primordial oscillations 
for a wide Euclid-like experiment covering the redshift 
range $0.9 \leq z \leq 1.8$ over a sky patch of 15,000 deg$^2$
around a fiducial value ${\cal A}_{\rm lin} = 0.03$ $\left({\cal A}_{\rm log} = 0.03\right)$ are 
$\sigma\left({\cal A}_X\right) \simeq 0.0025\ (0.0034)$ for $\log_{10}\left(\omega_X\right) = 0.1$,
$\sigma\left({\cal A}_X\right) \simeq 0.0017\ (0.0018)$ for $\log_{10}\left(\omega_X\right) = 1.1$,
$\sigma\left({\cal A}_X\right) \simeq 0.0041\ (0.0026)$ for $\log_{10}\left(\omega_X\right) = 2.1$
and for a deeper experiment as PFS  covering the redshift range $0.8 \leq z \leq 2.4$ over a sky 
patch of 1,464 deg$^2$ are
$\sigma\left({\cal A}_X\right) \simeq 0.0044\ (0.0032)$ for $\log_{10}\left(\omega_X\right) = 0.1$,
$\sigma\left({\cal A}_X\right) \simeq 0.0026\ (0.0029)$ for $\log_{10}\left(\omega_X\right) = 1.1$,
$\sigma\left({\cal A}_X\right) \simeq 0.0096\ (0.0064)$ for $\log_{10}\left(\omega_X\right) = 2.1$, 
in combination with {\em Planck}-like CMB temperature and polarization anisotropies and 
assuming $k_{\rm max} = 0.6\ h/$Mpc. We find an improvement by a factor 2 including 
non-linear scales from $k_{\rm max} = 0.1\ h/$Mpc to $k_{\rm max} = 0.6\ h/$Mpc.

Oscillatory features in the PPS also generate highly correlated signals in terms of 
non-Gaussianities \cite{Chen:2006xjb,Chen:2008wn,Flauger:2010ja,Chen:2010bka} and 
specific features appear also in the bispectrum (see Ref.~\cite{Chen:2010xka} for a review), 
so that primordial features can also be searched for in the bispectrum \cite{Akrami:2019izv}, 
or jointly in the power spectrum and bispectrum 
\cite{Fergusson:2014tza,Meerburg:2015owa,Karagiannis:2018jdt}. 
In addition, a scale-dependent contribution to the clustering bias is expected in the 
presence of primordial non-Gaussianity \cite{Dalal:2007cu,Matarrese:2008nc,Desjacques:2008vf,Slosar:2008hx}.
This last effect has been studied in Ref.~\cite{Ballardini:2017qwq} for large-scale features 
and in Ref.~\cite{Cabass:2018roz} for oscillatory features resulting in a very small effect 
that upcoming surveys will be unable to detect.
Direct studies of the imprint from these oscillatory features on the matter bispectrum 
are still promising in order to have a tighter and more robust detection of this 
features. A first investigation of the matter bispectrum in these models has been presented 
in Ref.~\cite{Vasudevan:2019ewf} using perturbation theory, and it will be  important to further 
extend this framework in order to see how non-linearities affect higher-order statistics.

\section*{Acknowledgements}
The authors are thankful to Hitoshi Murayama, Gabriele Parimbelli, Sergey Sibiryakov, and Zvonimir Vlah 
for helpful discussions and suggestions. We thank Matteo Biagetti and Xingang Chen for discussions 
and comments on the draft. The simulations have been performed on the Ulysses SISSA/ICTP supercomputer, 
and on the DiRAC DIaL cluster at Leicester University. 
The authors are supported by the INFN-InDark grant.
MB and FF acknowledge financial contribution from the agreement ASI/INAF n.  2018-23-HH.0 "Attivit\`a 
scientifica per la missione EUCLID – Fase D". MB acknowledges also partial support by the South African 
Radio Astronomy Observatory, which is a facility of the National Research Foundation, an agency of the 
Department of Science and Technology, and also by the Claude Leon Foundation. FF acknowledges financial 
support by ASI Grant 2016-24-H.0. MV acknowledges financial support from the agreement ASI-INAF n.2017-14-H.0.

%%%%%%%%%%%%%%%%%%%%%%%%%%%%%%%%%%%%%%%%%%%%%%%%%%%%%%%%%%%%%%%%%%%%%%%%%%%%%%%

%%%%%%%%%%%%%%%%%%%%%%%%%%%%%%%%%%%%%%%%%%%%%%%%%%%%%%%%%%%%%%%%%%%%%%%%%%%%%%%
\end{document}